\documentclass[11pt,fleqn]{article}

\usepackage{amsfonts,amssymb}
\usepackage[authoryear]{natbib}
\usepackage{amsmath}               
\usepackage{url}
\usepackage{authblk}
\usepackage{enumitem}
\usepackage{epstopdf}
\usepackage[ansinew]{inputenc}
\usepackage{multirow}
\usepackage{subfigure}
\usepackage{placeins}
\usepackage{footmisc}
\usepackage{rotating}
\usepackage{float}
\usepackage{xcolor}
\usepackage{array}
\usepackage{blkarray}
\usepackage{bbm}
\usepackage{booktabs}
\usepackage{setspace}
\usepackage{geometry}
\usepackage[hyperfootnotes=false]{hyperref}
\usepackage{lmodern}
\usepackage{xifthen}
\usepackage[title]{appendix}
\newtheorem{algorithm}{Algorithm}
\DeclareMathOperator*{\argmin}{arg~min}

\hypersetup{pdfborder = 0 0 0,%
    pdftitle = A statistical analysis of time trends in atmospheric ethane,%
    pdfauthor = {Marina Friedrich, Eric Beutner, Hanno Reuvers, Stephan Smeekes, Jean-Pierre Urbain, Whitney Bader, Bruno Franco, Bernard Lejeune, Emmanuel Mahieu},%
    setpagesize = false,%
    pdfpagelayout = SinglePage,%
    bookmarksopen = true,%
    bookmarksnumbered = true,%
    pdfstartview = Fit,%
    linkcolor = black,%
    anchorcolor = black,%
    filecolor = black,%
    citecolor = black,%
    menucolor = black,%
    urlcolor = blue,%
    colorlinks = true,%
}
\makeatletter
\renewcommand\d[1]{\mspace{6mu}\mathrm{d}#1\@ifnextchar\d{\mspace{-3mu}}{}}
\makeatother

\hypersetup{pdfborder = 0 0 0,%
}
\makeatletter
\renewcommand*{\@fnsymbol}[1]{\ifcase#1\or*\or $\dagger$\else\@arabic{\numexpr#1-2\relax}\fi}
\makeatother
\title{\sc A statistical analysis of time trends in atmospheric ethane \thanks{Corresponding author: Marina Friedrich, E-mail: \href{mailto:friedrich@pik-potsdam.de}{friedrich@pik-potsdam.de}. Support to the Li\`{e}ge team has been primarily provided by the F.R.S. - FNRS (Brussels) under Grant J.0147.18. Emmanuel Mahieu is a Research Associate with F.R.S. - FNRS. The vital support from the GAW-CH programme of MeteoSwiss is acknowledged. Mission expenses at the Jungfraujoch station were funded by the F\'{e}d\'{e}ration Wallonie-Bruxelles. We thank the International Foundation High Altitude Research Stations Jungfraujoch and Gornergrat (HFSJG, Bern) for supporting the facilities needed to perform the observations. W. Bader has received funding from the European Union's Horizon 2020 research and innovation programme under the Marie Sklodowska-Curie grant agreement no. 704951, and from the University of Toronto through a Faculty of Arts \& Science Postdoctoral Fellowship Award.}}
\author[a]{Marina Friedrich}
\author[b]{Eric Beutner}
\author[c]{Hanno Reuvers}
\author[d]{Stephan Smeekes} 
\author[d]{Jean-Pierre Urbain\footnote{Deceased on October 1, 2016}}
\author[e]{Whitney Bader}
\author[f]{Bruno Franco} 
\author[g]{Bernard Lejeune}
\author[g]{Emmanuel Mahieu}
\begin{small}
\affil[a]{\small Potsdam Institute for Climate Impact Research - Member of the Leibniz
Association}
\affil[b]{\small Vrije Universiteit Amsterdam - Department of Econometrics}
\affil[c]{\small Erasmus University Rotterdam - Department of Econometrics}
\affil[d]{\small Maastricht University - Department of Quantitative Economics}
\affil[e]{\small Agence Wallone de l'Air et du Climat (AWAC)}
\affil[f]{\small University of Li\`{e}ge - Institute of Astrophysics and Geophysics }
\affil[g]{\small Universit\'{e} Libre de Bruxelles - Faculty of Sciences}

\date{}
\end{small}
\begin{document}

\maketitle
\vspace{-13mm}
\begin{abstract}
Ethane is the most abundant non-methane hydrocarbon in the Earth's atmosphere and an important precursor of tropospheric ozone through various chemical pathways. Ethane is also an indirect greenhouse gas (global warming potential), influencing the atmospheric lifetime of methane through the consumption of the hydroxyl radical (OH). Understanding the development of trends and identifying trend reversals in atmospheric ethane is therefore crucial. Our dataset consists of four series of daily ethane columns obtained from ground-based FTIR measurements. As many other decadal time series, our data are characterized by autocorrelation, heteroskedasticity, and seasonal effects. Additionally, missing observations due to instrument failure or unfavorable measurement conditions are common in such series. The goal of this paper is therefore to analyze trends in atmospheric ethane with statistical tools that correctly address these data features. We present selected methods designed for the analysis of time trends and trend reversals. We consider bootstrap inference on broken linear trends and smoothly varying nonlinear trends. In particular, for the broken trend model, we propose a bootstrap method for inference on the break location and the corresponding changes in slope. For the smooth trend model we construct simultaneous confidence bands around the nonparametrically estimated trend. Our autoregressive wild bootstrap approach, combined with a seasonal filter, is able to handle all issues mentioned above.\footnote[1]{We provide R code for all proposed methods on \url{https://www.stephansmeekes.nl/code}.}\\
\end{abstract}


\numberwithin{equation}{section}
\section{Introduction}
There are several important reasons to study ethane time series. First, ethane is an indirect greenhouse gas influencing the atmospheric lifetime of methane. It degrades by reacting with the same oxidizer, the hydroxyl radical (OH; \citet{Aikin,Rudolf}), which is needed for the degradation of other major greenhouse gases such as methane. The OH radicals which are occupied by ethane are not available for the destruction of other pollutants \citep{Collins}. Second, ethane is an important precursor of tropospheric ozone \citep[see e.g.][]{Fischer,Franco2}. It contributes to the formation of ground level ozone (O$_3$) which is - unlike stratospheric ozone - a major pollutant affecting air quality. While ozone in higher levels of the atmosphere protects us from the Sun's harmful ultraviolet rays, ground level ozone damages ecosystems and has adverse effects on the human body. Third, ethane emissions can be used as a measure of methane emissions \citep[e.g.][]{Schaefer}. Both gases share some of their anthropogenic sources, while ethane does not have natural sources, methane is released in the atmosphere by both natural and anthropogenic activities. This makes it hard to measure the fraction of methane released by the oil and gas sector. An estimate of this fraction can be provided with the help of ethane measurements. Its monitoring is therefore crucial for the characterization of air quality the transport of tropospheric pollution. The main sources of ethane are located in the Northern Hemisphere, and the dominating emissions are associated to production and transport of natural gas \citep{Xiao}. 

Understanding recent and past developments in such emission data builds on the analysis of time trends. Trend estimation has received much attention in econometrics and statistics and many tools are available for this purpose. Trend estimation, however, is not enough; it is crucial to indicate the corresponding uncertainty around the estimate. This is commonly achieved by constructing confidence intervals which enable us to judge the significance of our results.

As many other climatological time series, measurements of atmospheric ethane display characteristics which complicate the analysis. In particular, calculation of uncertainty measures becomes increasingly difficult. These characteristics include strong seasonality, different degrees of variability (e.g. significant inter-annual changes), and missing observations due to instrument failures or unfavorable measurement conditions. Atmospheric ethane, when measured with the Fourier Transform InfraRed (FTIR) remote-sensing technique, is a prominent example in which all three problematic characteristics arise. It displays strong seasonality, a time-varying variance and, since measurements can only be taken under clear sky conditions, many missing data points. Therefore, it is important to use methods which provide reliable results under these circumstances.

Bootstrap methods can address some of these problems, as in \citet{Gardiner}. The authors propose a method for (linear) trend analysis of greenhouse gases. \citet{Gardiner} stress that the residuals of the model are serially correlated and not normally distributed. They propose an i.i.d. (independently and identically distributed) bootstrap method to construct confidence intervals around the slope parameter. This approach suffers from two major drawbacks. First, the approach does not provide confidence intervals for the break location. Second, in the presence of autocorrelation, the i.i.d.~bootstrap method cannot correctly mimic the dependence structure of the residuals. Alternative bootstrap methods, such as the block or sieve bootstrap, are available to solve this problem. In terms of implementation, both require only minor modifications compared to the i.i.d.~bootstrap.

Similar methods as in \citet{Gardiner} have been applied to various data series. \citet{DeSmedt2010} investigate trends in satellite observations of formaldehyde columns in the troposphere, \citet{Nog2011} study linear trends in ice phenology data, and \citet{Mahieu2014} look at stratospheric hydrogen chloride increases in the Northern Hemisphere. More recently, \citet{Hausmann2016} use a bootstrap method to study trends in atmospheric methane and ethane emissions measured at Zugspitze and Lauder. The latter two papers split the sample into two periods and compare the changes in trends. It is, however, not always obvious where the sample should be split, and user-selected break points are thus somewhat arbitrary. This issue can be resolved using data-driven methods to select the break point. While trend estimates such as slopes for linear approaches, usually come with confidence intervals, the break location is often stated without any measure of uncertainty. Obtaining confidence intervals for break locations gives valuable additional insights.

This paper aims to analyze trends in atmospheric ethane with an alternative set of statistical tools. Our dataset consists of four series of daily ethane columns (i.e.~the number of molecules integrated between the ground and top of the atmosphere in a column of a given area, e.g. a square centimeter) obtained from ground-based FTIR measurements. We present selected methods designed for the analysis of time trends and trend reversals and apply them to our dataset. We focus on two different, but complimentary approaches which are particularly suited in this context. First, we present a linear trend model which allows for a break at an unknown time point for which we also obtain confidence intervals. It provides researchers with a tool to test for the presence of a break and, if so, it additionally gives an estimate of its location together with a reliable confidence interval. With this method, it is not necessary to split the sample. If there is a break present, it automatically determines two different estimates of the trend parameters.

In the second part, we move to a more flexible specification by considering a smoothly varying nonparametric trend model. It does not impose any assumptions regarding the form of the trend. However, the trend function will no longer be characterized by merely two values - like the intercept and slope for a linear trend. The nonparametric approach results in a collection of estimates, one for every time point, which together define the trend. We are nevertheless concerned with investigating certain properties of the resulting trend. This is why we propose three additional methods to take a closer look at the trend shape. 

In both parts, we suggest the use of bootstrap methods to construct confidence intervals and obtain critical values for our statistical tests. We advocate the use of a specific bootstrap method - the autoregressive wild bootstrap - which is applicable to correlated and heteroskedastic data. Its second advantage over many other bootstrap methods is that it can easily be applied to data series with missing observations. 

The structure of the paper is as follows. The data description and general modeling approach are introduced in Section 2. Section 3 presents the linear trend approach and corresponding ethane results. Section 4 continues with the nonparametric trend model. The first part gives the model specifications and discusses the results. In the second part, we present how to conduct inference on the shape of the nonparametric trends and apply these methods to the data. Section 5 concludes. In the supplementary appendix, we give additional technical details in part A and provide a Monte Carlo simulation study in part B.

\section{Trends in atmospheric ethane}

\subsection{The data}
We study four series of atmospheric ethane measurements. The measurement stations are located at Jungfraujoch (Swiss Alps), Lauder (New Zealand), Thule (Greenland), and Toronto (Canada). Jungfraujoch, Thule and Toronto lie in the Northern Hemisphere while Lauder is located in the Southern Hemisphere. 

The Jungfraujoch measurement station is located on the saddle between the Jungfrau and the M\"onch, at 46.55$^{\circ}$ N, 7.98$^{\circ}$ E, 3580 m altitude \citep{Zander}. The time series consists of daily ethane columns recorded under clear-sky conditions between 1986 and 2019 with a total of 2935 data points. Part of the series, from 1994 to 2014, has been analyzed in \citet{Franco} and \citet{FSU}. It is an interesting series to study, since the measurement conditions are very favorable at this location due to high dryness and low local pollution. This is the longest FTIR time series of ethane, with more than three decades of continuous measurements available. Further details on the ground-based station at Jungfraujoch and on how measurements are obtained can be found in \citet{Franco}. 

The Lauder time series starts in 1992 and ends in 2014 and has 2550 observations. The station is located at 45$^{\circ}$ S, 170$^{\circ}$ E, 370 m altitude. A part of the series (until 2009) was investigated in \citet{Zeng}. The measurement station in Thule is located at 76.52$^{\circ}$ N, 68.77$^{\circ}$ W, 225 m altitude. The series consists of 814 data points taken between 1999 and 2014. Finally, the Toronto station is located at 43.66$^{\circ}$ N, 79.40$^{\circ}$ W, 174 m altitude and the series ranges from 2002 to 2014 with 1399 observations. The series obtained at Thule and Toronto have been studied in \citet{Franco2}. Whenever multiple measurements are taken on one day, a daily mean is considered. 

The Jungfraujoch series contains an average of 89.9 data points per year, corresponding to data availability of about 25\% on yearly basis, Lauder has on average 115.9 data points per year (32\%), Thule 54.4 (15\%) and Toronto 112.1 (31\%). These percentages clearly indicate that missing data is a severe and non-negligible problem in this type of analysis. In particular, simple imputation is likely to be imprecise and it may introduce strong biases into the outcomes. We therefore do not impute the data but use statistical methods that directly allow for missing values.

These data are also characterized by a strong seasonal pattern, as ethane degrades faster under warm weather conditions than in cold temperature and therefore, the measurements display local peaks every winter period. Finally, it is worthwhile to note that the measurements display strong autocorrelation, which has to be accommodated as well. 

\subsection{A general trend model}
\label{sec:model}
Let $y_t$ denote ethane measurements at time $t$, where $t$ ranges from 1 to $T$. Then we formulate the trend model
\begin{equation} \label{eq:genmodel}
y_t = d_t + s_t + u_t,
\end{equation}
where $d_t$ is the long-run trend -- our object of interest, $s_t$ is the (deterministic) intra-annual seasonal pattern, and $u_t$ is a stochastic error term that captures short-run fluctuations. We assume that $u_t$ is of the form $\sigma_t v_t$ where $\sigma_t$ is a deterministic sequence and $v_t$ is a linear process with absolutely summable coefficients. Thus, $(u_t)$ can exhibit heteroskedasticity and serial dependency, as also observed in our ethane measurement series. While this structure implies that dependence dies out over time, this can be quite slow and therefore fairly strong autocorrelation is allowed for. Moreover, it is well known that the autoregressive wild bootstrap is valid for many problems with this error specification.  

Seasonal effects are modeled through $s_t$; we focus on a deterministic specification given the fixed nature of seasonal effects in this context, though stochastic effects can be allowed for as well. We model and estimate the seasonal effects with the help of Fourier terms:
\begin{equation}
s_t=\sum_{j=1}^S a_j\cos(2j\pi t)+b_j\sin(2j\pi t).
\label{eq:Fourier}
\end{equation}
This specification of the seasonal variability is widely used when estimating trends in atmospheric gases, see e.g. \citet{Gardiner}, \citet{Franco}, and \citet{Franco2}. These papers show that the variability is well captured by the inclusion of $S=3$ sine and cosine terms. Our own investigations confirm that three terms capture the seasonal variation well,\footnote{Detailed results are available on request.} and therefore we follow the same approach and consider equation \eqref{eq:Fourier} with $S=3$ in the remainder of the paper.

The specification of $d_t$ depends on our trend specifications, see Sections \ref{sec:parametric} and \ref{sec:nonpara}. Before going into detail, we first address the missing data issue. Define the binary variable $M_t$ as 
\begin{equation}
M_t = \left\{
\begin{array}{ll}
1 & \text{if $y_t$ is observed} \\
0& \text{if $y_t$ is missing}
\end{array}
\qquad t = 1, \ldots, T. \right.
\label{eq:missings}
\end{equation}
In order to derive theoretical properties of our methods, e.g. \citet{FSU}, one typically assumes that the missing pattern, characterized by $\{M_t\}$, is independent of the observations. Strictly speaking, in the present case we cannot exclude a mild dependence between ethane and the missing pattern as ethane's primary sink is oxidation by the hydroxyl radical, which is dependent on solar insolation.\footnote{We thank an anonymous referee for pointing this out.} However, given the atmospheric lifetime of ethane in relation to our sampling frequency, we argue that any dependence is negligible in comparison to other fluctuations since our purpose is analyzing long-term trends. Ethane's lifetime is of the order of two months, while FTIR measurements are taken on average every three to four days. As such, the high frequency of measurements means that most variation in ethane that we capture is due to other sources. 

We allow the probability of observing measurements on a given day to vary over time, which can accommodate for instance seasonal variation and long-term climatic trends. In addition, we assume that the probability of observing a measurement on a given day may be serially dependent, but we need this dependence to decay over time; for the precise meaning we have in mind please see \cite{FSU}, Assumption 4. This ensures that we, over a large enough time span, always have sufficient data available to estimate the trend. It is reasonable to assume that the pattern of the missing data points in the case of FTIR measurements -- generally caused by adverse weather conditions -- satisfy these assumptions.

A second source of missing data -- resulting in prolonged periods without observations -- might be instrument failure and/or maintenance, as well as polar nights for stations close to the poles. While instrument failure, if it is not expected to last indefinitely, can be captured by an assumption like Assumption 4 in \cite{FSU} and polar nights can be modeled by varying the probability of missing data, we stress that for prolonged periods without observations, one cannot draw meaningful conclusions. Practically one needs data around the point of interest to estimate the trend and conduct inference. While for the linear approach such periods are less of an issue as long as the break in trend is not thought to be located in such a period, the nonparametric approach in Section 4, which requires to construct local averages around the date, becomes completely uninformative. Such periods should therefore be treated with caution, and would have be excluded from the analysis in order to draw meaningful conclusions. The reader is referred to \citet{FSU} for a more precise statement and detailed discussion of these assumptions. 

\section{Modeling trends linearly}
\label{sec:parametric}
\subsection{A broken trend model}
We now return to the trend model of \eqref{eq:genmodel} and specify $d_t$ as follows:
\begin{equation}
d_t=\alpha+\beta t+\delta D_{t,T_1},
\label{model}
\end{equation}
where
\begin{equation}
D_{t,T_1}=\begin{cases} 
0   &   \mbox{if } t\leq T_1, \\ 
t-T_1  &   \mbox{if } t>T_1.\\
\end{cases}
\label{timedummy}
\end{equation}

Equations \eqref{model} and \eqref{timedummy} describe a broken linear trend model with a single\footnote{\cite{BaiPerron} discuss inference in regression models with multiple unknown breaks. One of their findings is that break locations can be estimated sequentially. An extension of our bootstrap methodology to multiple structural changes is left for future research.} and unknown break at date $T_1$. The intercept and slope parameter before the break are $\alpha$ and $\beta$, respectively. For $t>T_1$, the dummy variable $D_{t,T_1}$ induces a change in the slope coefficient from $\beta$ to $\left(\beta+\delta\right)$ while altering the intercept in such a way as to enforce continuity at the break date. This prevents the modeled ethane concentration from exhibiting a sudden unrealistic jump at $T_1$.

The parameters of interest are $(\alpha,\beta,\delta)$, the parameters in the Fourier specification \eqref{eq:Fourier}, and the unknown breakdate $T_1$. For future reference we denote the fitted seasonal effects by $\hat{s}_t$. The inherent simplicity and small number of parameters make \eqref{model} easy to estimate and interpret. Both aspects make linear trend models a popular tool for trend analysis \citep[see e.g.][]{Bloomfield,FV,MV}. However, one should realize that piecewise linearity is most likely nothing but an approximation of reality. As such, we view the broken trend model as a description of the most prominent trend features and designate any remaining nonlinearities to the error term.

\subsection{Testing for a break}
Following \citet{BaiPerron}, we propose a formal test to determine whether a model with one break is preferred over a simple linear trend model. Let $\Lambda$ denote the set of possible break dates. For some $0<\lambda<1/2$, we specify this set as $\Lambda=\left[\lambda T,(1-\lambda)T\right]$, that is, we require the break date to be bounded away from the boundaries of the sample. This assumption is standard in the structural breaks literature. Without this assumption the test statistic will diverge as $T\to\infty$ and the method will not have any asymptotic validity (see section 5.2 of \citet{Andrews1993} for details). In practice, $\lambda$ has to be specified by the user. Its choice should ensure that sufficient data points are available on both sides of each candidate break to allow for the estimation of the unknown parameters. We set $\lambda=0.1$. Changes in $\lambda$ have little effect on the results as long as the estimated break point does not occur too close to the boundaries of $\Lambda$.\footnote{We thank an anonymous referee for pointing out the difficulties that can occur when choosing $\lambda$ in practice. In general, the practitioner should proceed with care when the estimated break date is in close proximity to the start or end of the sample. Re-estimating the model with a slightly different value of $\lambda$ should indicate whether results should be treated with caution.} Empirical evidence for this claim can be found in Table \ref{tab:varylambda}. As visible in this table, changes in $\lambda$ lead to qualitatively similar confidence intervals.

\begin{table}[t]
\centering
\begin{tabular}{lccccc}
\noalign{\smallskip} \hline \hline 
 & $\lambda=5\%$ & $\lambda=10\%$ & $\lambda=15\%$  \\
\hline
\noalign{\smallskip}
Jungfraujoch & [2005.59,2007.19] & [2005.66,2007.04] & - \\
Lauder   & [1996.37,2009.65]            & [1996.60,2008.85]       & [1997.12,2008.72] \\
Thule    & [2005.17,2009.28]            & [2005.22,2009.58]       & [2005.21,2010.21] \\
Toronto  &  [2008.26,2009.71]           & [2008.06,2010.04]        & [2008.16,2009.66]             \\
\noalign{\smallskip} \hline \noalign{\smallskip}
\end{tabular}
\caption{The confidence intervals for the break date for various choices of the trimming parameter $\lambda$.} 
\label{tab:varylambda}
\end{table}

Having specified $\Lambda$, we define our test statistic as
\begin{equation}
F_T=\min_{\alpha,\beta,s_t}\sum_{t=1}^{T}M_t\left(y_t-\alpha-\beta t-s_{t}\right)^2-\underset{T_c\in \Lambda}{\operatorname{inf}}\min_{\alpha,\beta,\delta,s_t}\sum_{t=1}^{T}M_t\left(y_t-\alpha-\beta t-\delta D_{t,T_c}-s_{t}\right)^2,
\label{eq:break_stat}
\end{equation}
where we compare the sum of squared residuals of a model without break to the lowest sum of squared residuals of a model including one break. It is a formal test of the pair of hypothesis $H_0:\delta=0$ versus $H_1:\delta\neq0$, for every possible break point $T_c\in\Lambda$. Low (high) values of $F_T$ indicate little (substantial) evidence in favor of the model with a structural break. Given a significance level of the test, the critical value of the test determines the cut-off point. The exact procedure is summarized below.

\begin{algorithm}[Autoregressive Wild Bootstrap - Break test]
$\phantom{1}$
\begin{enumerate}
\item Calculate residuals from the estimation of model \eqref{eq:genmodel} with the trend $d_t$ specified by \eqref{model}, with $\delta=0$. For $t=1,...,T$, \begin{equation*}\hat{u}_t=M_t\left(y_t-\hat{\alpha}-\hat{\beta}t-\hat{s}_t\right).\end{equation*} 
\item For $0 < \gamma < 1$, generate $\nu_1^\ast,\ldots,\nu_n^\ast$ as i.i.d.~$\mathcal{N}(0,1-\gamma^2)$ and let $\xi_t^* = \gamma \xi_{t-1}^* + \nu_t^*$ for $t=2,\ldots,T$. Take $\xi_1^* \sim \mathcal{N}(0,1)$ to ensure stationarity of $\{\xi_t^*\}$.
\item Calculate the bootstrap errors $u_t^{\ast}=M_t\xi_t^{\ast}\hat{u}_t$ and generate the bootstrap sample as \begin{equation*}
y_t^\ast=M_t\left(\hat{\alpha}+\hat{\beta}t+\hat{s}_t+u_t^\ast\right)
\end{equation*} for $t=1,...,T$, using the same estimated coefficients as in Step 1.
\item Obtain $F_T^{\ast}$ from $y_t^{\ast}$ as in equation \eqref{eq:break_stat} and store the result.
\item Repeat Steps 2 to 4 $B$ times to obtain the bootstrap distribution of $F_T^{\ast}$.
\end{enumerate}
\end{algorithm}
Since the test is rejected for large values of the test statistic $F_T$, we use the $(1-\alpha)$ quantile of the ordered bootstrap statistics as critical value for the break test. In Step 2 of the above algorithm, the autoregressive coefficient $\gamma$ has to be chosen. The choice reflects a trade-off: a larger value captures more of the dependence whereas a smaller value allows for more variation in the bootstrap samples. We suggest to follow \citet{FSU} and use $\gamma=\theta^{1/l}$ with $l=1.75T^{1/3}$ and $\theta=0.1$.


In Step 2 we generate $\left\{\xi_t^{\ast}\right\}$ for all $t=1,\ldots,T$, although in Step 3 we construct bootstrap errors and subsequently, bootstrap observations only when there exists an actual data point. This is what the multiplication by $M_t$ in Step 3 ensures. The bootstrap sample thus correctly reflects the missing pattern present in the data. 

The autoregressive wild bootstrap (AWB) can also be used to obtain confidence intervals for the unknown break date $T_1$ and all parameter estimates. We refer the reader to Appendix A of the supplementary material for further details.

\subsection{Empirical findings for ethane series}
\label{sec:app_para}
Panel (a) of Table \ref{tab:breaks} summarizes the results of the break test for the four ethane time series. As an example, for the Jungfraujoch, the test statistic of the F-test is $1.40\times10^{33}$, while the bootstrapped critical value lies at $5.54\times10^{31}$. The resulting p-value of $0$ indicates that the null hypothesis of no break should be rejected. The conclusions are similar for Lauder, Thule, and Toronto. We thus thus include a break point in each trend specification.

\begin{figure}[ht]
     \begin{center}
        \subfigure[Jungfraujoch]{
            \includegraphics[width=0.48\textwidth, clip, trim = {0.2cm 1cm 0cm 2cm}]{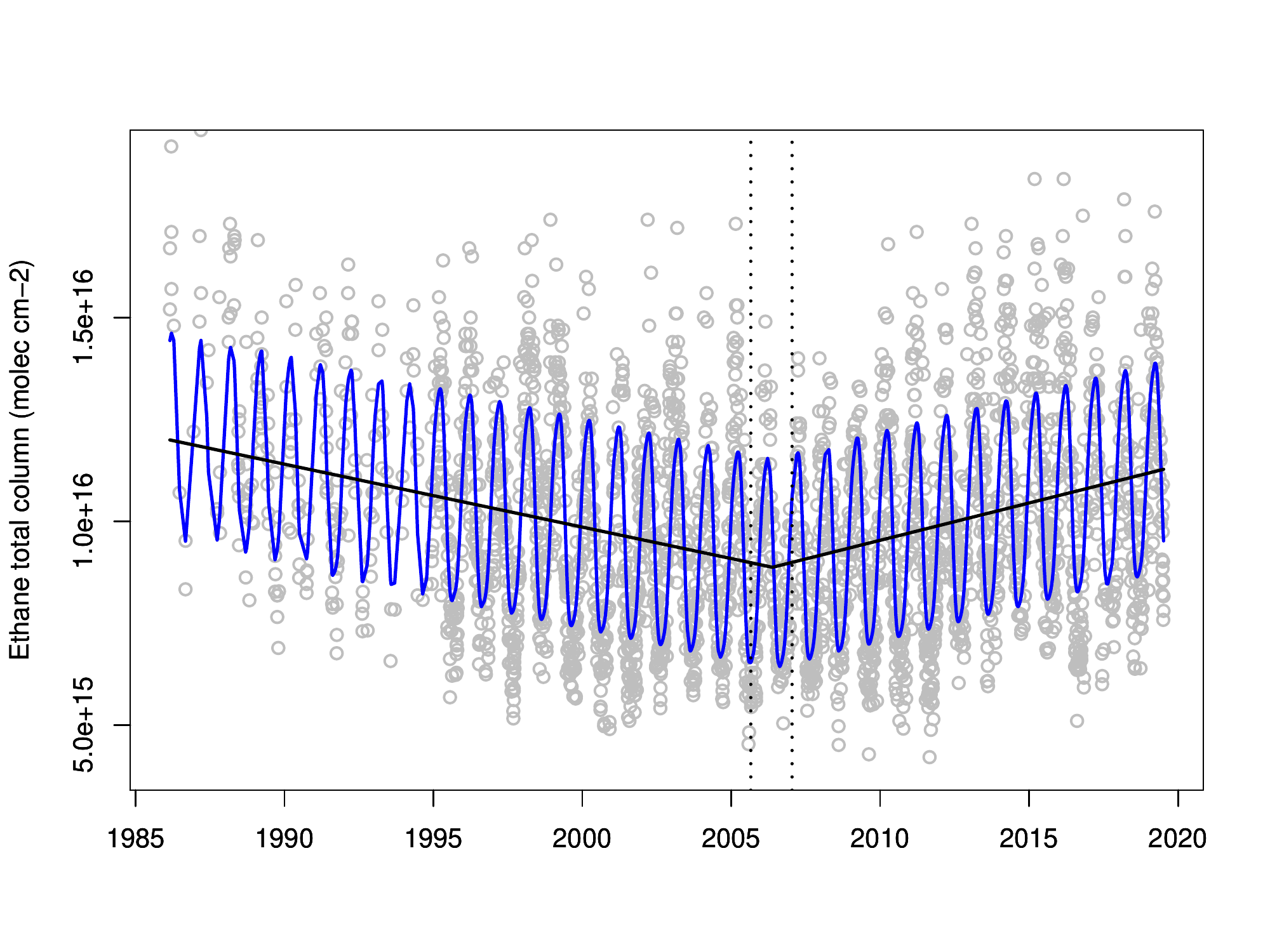}
        }
        \subfigure[Lauder]{
            \includegraphics[width=0.48\textwidth, clip, trim = {0.3cm 1cm 0cm 2cm}]{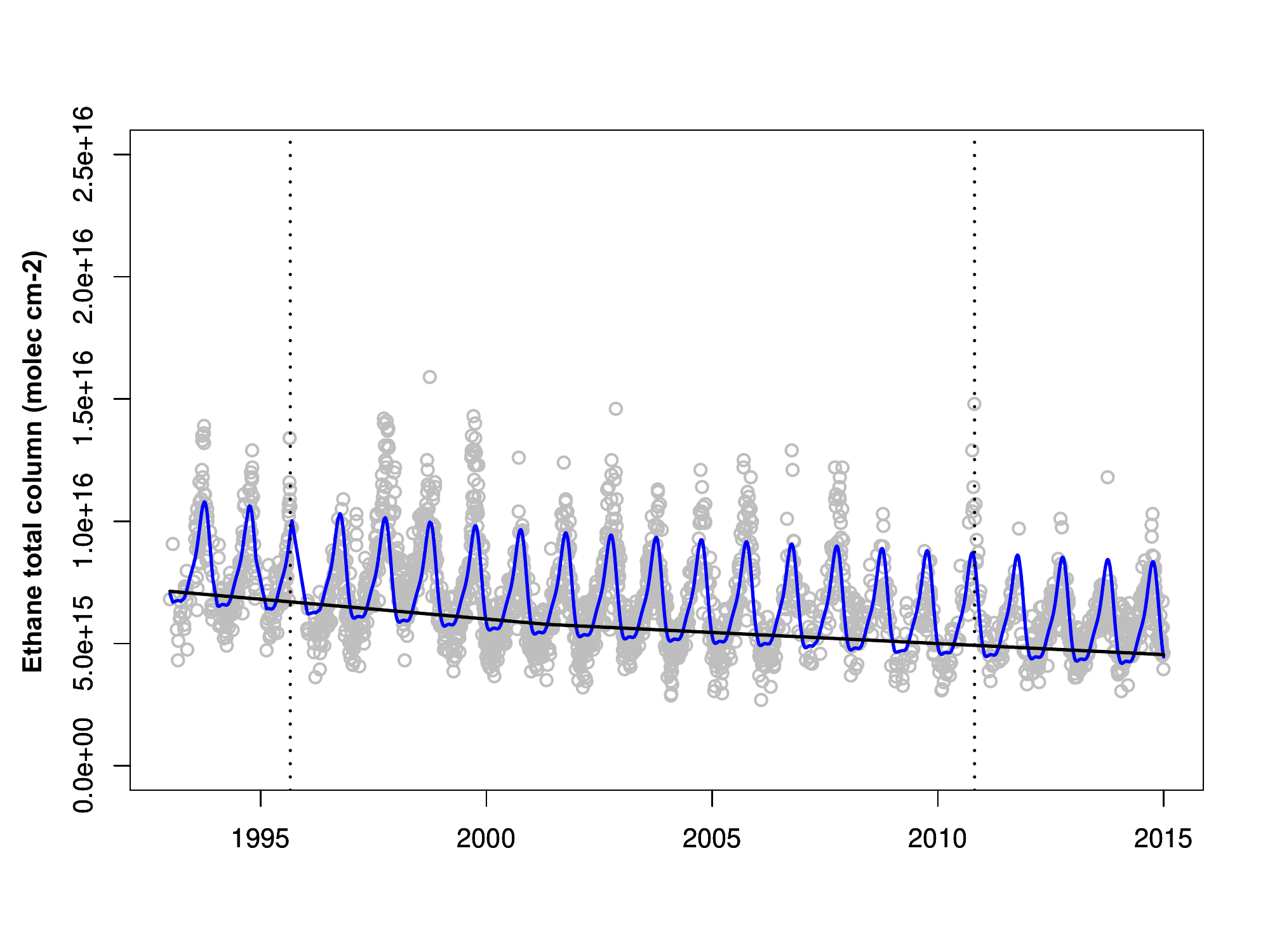}
        }\\
				  \subfigure[Thule]{
            \includegraphics[width=0.48\textwidth, clip, trim = {0.2cm 1cm 0cm 2cm}]{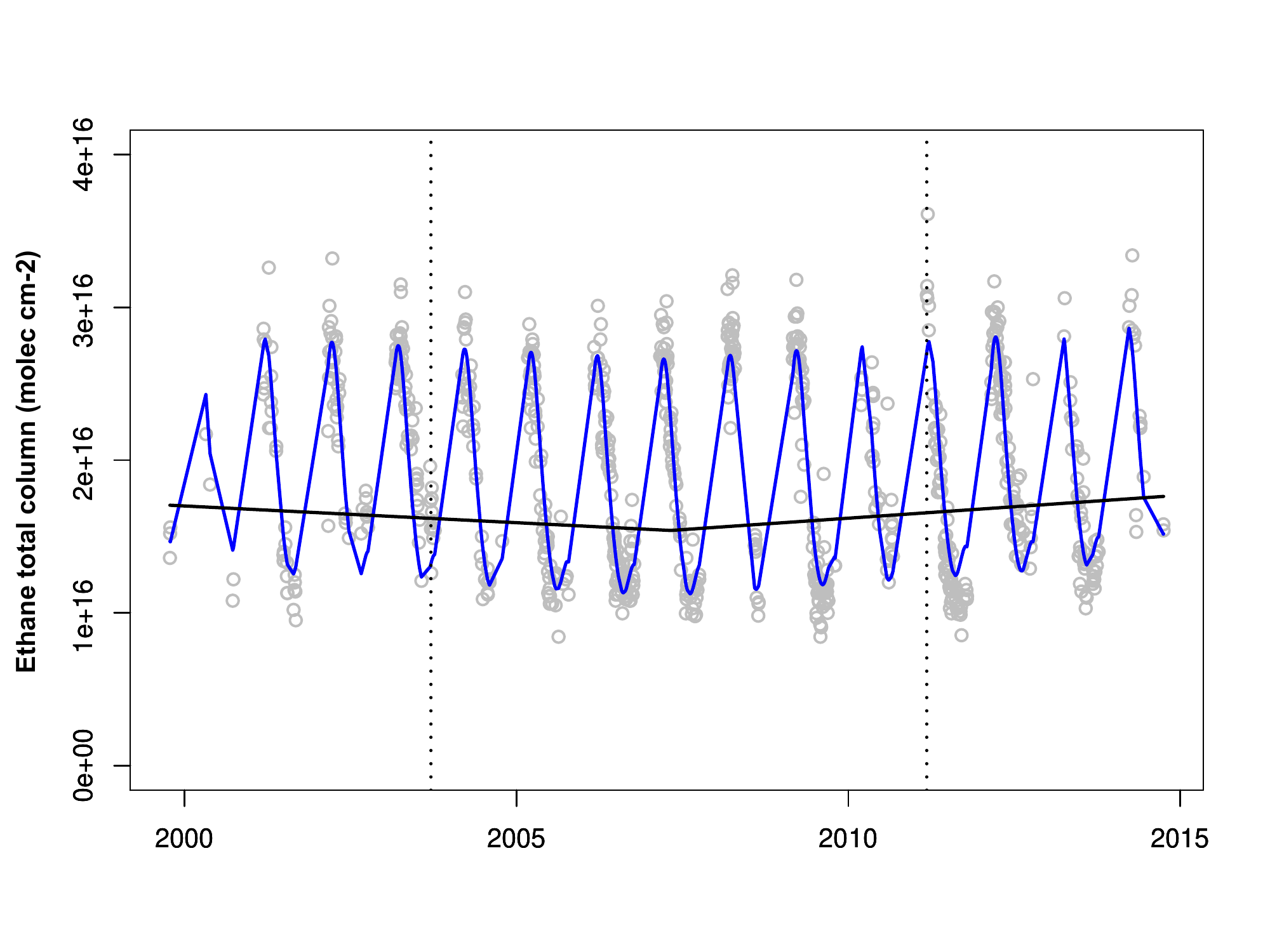}
        }
        \subfigure[Toronto]{
            \includegraphics[width=0.48\textwidth, clip, trim = {0.3cm 1cm 0cm 2cm}]{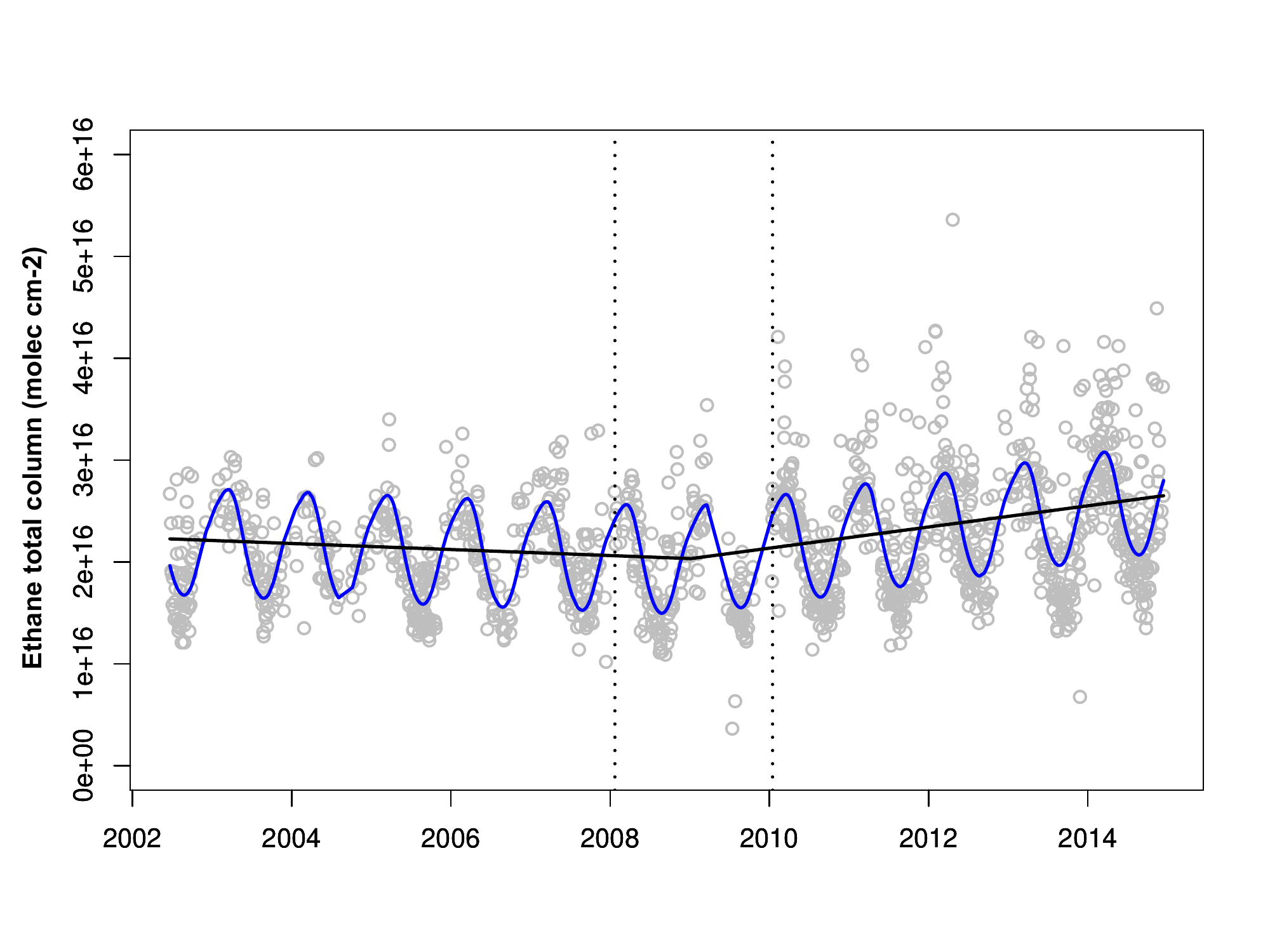}
        }
    \end{center}
    \caption{
        This figure shows the data (gray circles) as well as the continuous broken trend (black) and the fitted Fourier series (blue) for all four series.
     }
   \label{fig:parametric}
\end{figure}

The estimated break location for the Jungfraujoch series is 2006.38 (19.05.2006) and the AWB method provides a confidence interval ranging from 2005.66 to 2007.04 (26.08.2005 to 14.01.2007). The graphical summary in Figure \ref{fig:parametric}(a) plots: the ethane time series (gray circles), the seasonal fit of three Fourier terms (blue), the estimated broken trend (black), and the confidence interval of the break date (dotted vertical lines). We observe a significant decrease in ethane concentration of about $-1.54\times 10^{14}$ mol $\text{cm}^{-2}$ $\text{yr}^{-1}$ before the break, followed by an increase of $1.83\times 10^{14}$ mol $\text{cm}^{-2}$ $\text{yr}^{-1}$ after the break. Figures \ref{fig:parametric}(b)-(d) and Panel (B) of Table \ref{tab:breaks} provide information on the other series.

\begin{table}[t]
\centering
\begin{tabular}{llllll}
\noalign{\smallskip} \hline \hline 
\multicolumn{6}{c}{\textbf{A: Break test results}} \\
\noalign{\smallskip} \hline
 & $T$ & Sample period &  \textit{p}-value & $S_T$ & Critical value  \\
\hline
\noalign{\smallskip}
Jungfraujoch & 2935 & 1986-2019 & $0.0000$ & $1.40\times10^{33}$ & $5.54\times10^{31}$ \\
Lauder & 2550 & 1992-2014 &$0.0248$ & $2.82\times10^{31}$ & $2.30\times10^{31}$ \\
Thule & 814 & 1999-2014 &$0.0000$ & $1.99\times10^{32}$ & $0.75\times10^{32}$ \\
Toronto & 1399 & 2002-2014&$0.0000$ & $1.93\times10^{33}$ & $2.84\times10^{32}$ \\
\noalign{\smallskip} \hline
\multicolumn{6}{c}{\textbf{B: Break dates and parameter estimates}} \\
\noalign{\smallskip} \hline
& Break & [CI]& \textit{period} & Slope & [CI] \\
\hline
\noalign{\smallskip}
\multirow{2}{*}{Jungfraujoch}	& \multirow{2}{*}{2006.38} & \multirow{2}{*}{[2005.48,2007.39]}	&\textit{before} 	&$-1.54\times10^{14}$ & [$-1.74\times10^{14}$,$-1.36\times10^{14}$] \\
														&	& &\textit{after}	&$\textcolor[rgb]{1,1,1}{-}1.83\times10^{14}$ & [$\textcolor[rgb]{1,1,1}{-}1.58\times10^{14}$,$\textcolor[rgb]{1,1,1}{-}2.05\times10^{14}$] \\[0.3cm]
\multirow{2}{*}{Lauder}		& \multirow{2}{*}{2001.34} & \multirow{2}{*}{[1992.33,2007.03]} 	&\textit{before}	&$-1.62\times10^{14}$ & [$-1.97\times10^{14}$,$-1.26\times10^{14}$] \\
														&	& &\textit{after}	&$-9.06\times10^{13}$ & [$-1.08\times10^{14}$,$-7.26\times10^{13}$] \\[0.3cm]
\multirow{2}{*}{Thule}		& \multirow{2}{*}{2007.32} & \multirow{2}{*}{[2003.99,2010.94]} 	&\textit{before}	&$-2.19\times10^{14}$ & [$-3.51\times10^{14}$,$-8.68\times10^{13}$] \\
													&		& &\textit{after}	&$\textcolor[rgb]{1,1,1}{-}3.00\times10^{14}$ & [$\textcolor[rgb]{1,1,1}{-}1.89\times10^{14}$,$\textcolor[rgb]{1,1,1}{-}4.14\times10^{14}$] \\[0.3cm]
\multirow{2}{*}{Toronto}		& \multirow{2}{*}{2008.96} & \multirow{2}{*}{[2008.12,2009.87]}	&\textit{before}	&$-2.96\times10^{14}$ & [$-4.51\times10^{14}$,$-1.40\times10^{14}$] \\
													&		& &\textit{after}	&$\textcolor[rgb]{1,1,1}{-}1.04\times10^{15}$ & [$\textcolor[rgb]{1,1,1}{-}8.64\times10^{14}$,$\textcolor[rgb]{1,1,1}{-}1.20\times10^{15}$] \\
\noalign{\smallskip} \hline \noalign{\smallskip}
\end{tabular}
\caption{Panel (A): Sample period and sample size $T$, as well as results of the break tests (with $\lambda=0.1$): $p$-value, test statistic $S_T$ (as in eq. (3.5)) and bootstrap critical value obtained as in Algorithm 1. Panel (B): Point estimate and confidence interval [CI] of the break date $T_1$ as well as the slope parameter $\beta$ (in mol $\text{cm}^{-2}$ $\text{yr}^{-1}$) before and after the break.}
\label{tab:breaks}
\end{table}

Our results are qualitatively similar to those in \citet{Franco}.\footnote{A difference with the results in \citet{Franco} is the estimated break data. Based on data until August 2014, they place the break point beginning 2009. Two facts can explain this apparent discrepancy. First, the break point in \citet{Franco} was determined by finding the minimum in the running mean of the daily average data instead of selecting the break point that achieves the minimum sum of squared residuals among all candidate broken trend models. Second, \citet{Franco} do not report a confidence interval for their break location as appropriate bootstrap methodology was not available at the time. As such it is hard to judge whether our outcomes are significantly different.} As mentioned there, the initial downward trend can be explained by general emission reductions since the mid 1980's of the fossil fuel sources in the Northern Hemisphere. This has also been reported by other studies. The upward trend seems to be a more recent phenomenon. Some studies attribute it to the recent growth in exploitation of shale gas and tight oil reservoirs, taking place in North America, see e.g.~\citet{Vinci}, \citet{Franco2} and \citet{Helmig2016}. The significant negative coefficients before and after the break in Panel (B) of Table \ref{tab:breaks} indicate that Lauder is not yet impacted by the recent increase of ethane in the Northern Hemisphere, further from the main emission sources. Lauder is the only site in the data set which is located in the Southern Hemisphere. Indeed, C$_2$H$_6$ has a mean atmospheric lifetime of 2 months, significantly shorter than the time needed to mix air between both hemispheres \citep{Simpson}.

\section{Modeling trends as smooth nonparametric functions}
\label{sec:nonpara}
The piecewise linear model provides a transparent overview of the long-term behavior of the ethane concentration. That is, fitted trends (as seen in Figure \ref{fig:parametric}) provide a clear visualisation of periods of decreasing/increasing ethane concentration. However, all short-lived deviations from this linear trend are not discernible. We will now introduce a more flexible the model which does not require functional form, comment on our empirical findings, and propose some tests that allow for a more detailed analysis of the data.

\subsection{The nonparametric trend model}
Instead of a linear specification of the trend $d_t$ in \eqref{eq:genmodel}, we now specify
\begin{equation}
d_t=g(t/T), \qquad t=1,...,T,
\label{eq:model_nonpara}
\end{equation}
where $g(\cdot)$ denotes a smooth (i.e.~twice-differentiable) function defined on the interval $[0,1]$. As is standard with this approach \citep[see e.g.][]{Robinson,WuZhao}, we map all time points into the interval $\left[0,1\right]$ by the division by $T$, with the idea that when the sample size $T$ increases we observe points on a denser grid of $[0,1]$. This is mainly done for theoretical purposes, and does not affect estimation in practice.

The main goal is to estimate the function $g(\cdot)$ and determine the uncertainty around this estimate. We use the nonparametric kernel estimator suggested by \citet{Nadaraya} and \citet{Watson} in a two-stage procedure where we first eliminate seasonal variability and next estimate the trend function nonparametrically. The estimator uses a smoothing parameter called the bandwidth. Essentially, the bandwidth determines how many data points around the point of interest are used to estimate the trend by constructing a local (weighted) average around that point. Large bandwidths produce very smooth estimates, while for small bandwidth, estimated trends fluctuate more. Bandwidth selection is  important for this type of estimation \citep[e.g.][]{Fan}. If the bandwidth is too small, approaching zero, the trend estimate almost coincides with the data points, which would be overfitting. If, in contrast, the bandwidth is very large, the trend estimate will be close to a linear trend. An appropriate bandwidth lies in between to avoid over- or underfitting and ultimately has to be selected by the user. The choice depends on the context of the study. Data-based procedures exist which can help with bandwidth selection. However, it is not uncommon to encounter problems with these methods in practice. We elaborate on this in the next section.

This model was studied by \citet{FSU}, who develop bootstrap methods to construct confidence bands around the trend and establish the method's theoretical properties. Inference on the trend is conducted using the autoregressive wild bootstrap to construct pointwise intervals in a similar fashion as above. Subsequently, we apply a three step procedure to find simultaneous confidence bands based on the pointwise intervals. Many interesting research questions, like whether a coefficient remains zero over the whole period or whether there was an upward trend over a certain period of time, cannot be answered with pointwise confidence intervals. Therefore, we use simultaneous confidence bands as discussed in \citet{HM}, \citet{Buhlmann}, and \citet{NP}. For technical details on the estimation and bootstrap methods, and how to obtain simultaneous confidence bands, we refer to Appendix A of the supplementary material.  

\subsubsection{Smooth trends in ethane}
\label{sec:app_nonpara}
To estimate the trend function, we first obtain residuals from a regression of the ethane data on three Fourier terms. From these residuals we estimate the trend function nonparametrically using a local constant kernel estimator with an Epanechnikov kernel.\footnote{Other estimators, such as the local linear estimator, can be used as well \citep{Fan2,FG}. Other kernels can be used instead of the Epanechnikov; we find that results are insensitive to this choice. Details are available on request.} We illustrate a data-driven bandwidth selection using the Modified Cross Validation (MCV) approach of \citet{ChuMarron} which is discussed in the nonparametric trend setting in \citet{FSU}. Technical details can again be found in Appendix A.

\begin{figure}[th]
	\begin{center}
		\subfigure[Jungfraujoch]{
            \includegraphics[width=0.45\textwidth]{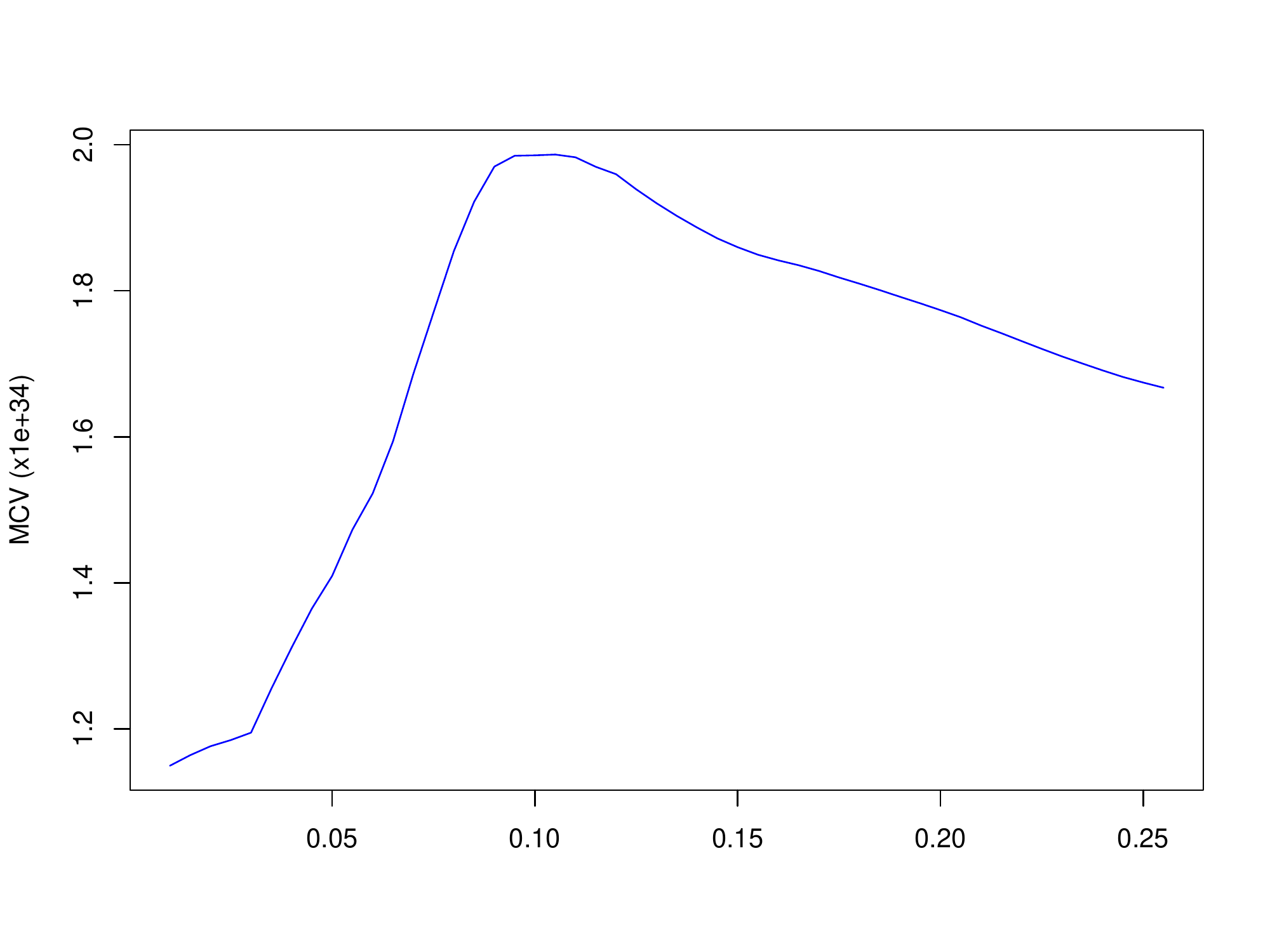}
        }
        \subfigure[Lauder]{
            \includegraphics[width=0.45\textwidth]{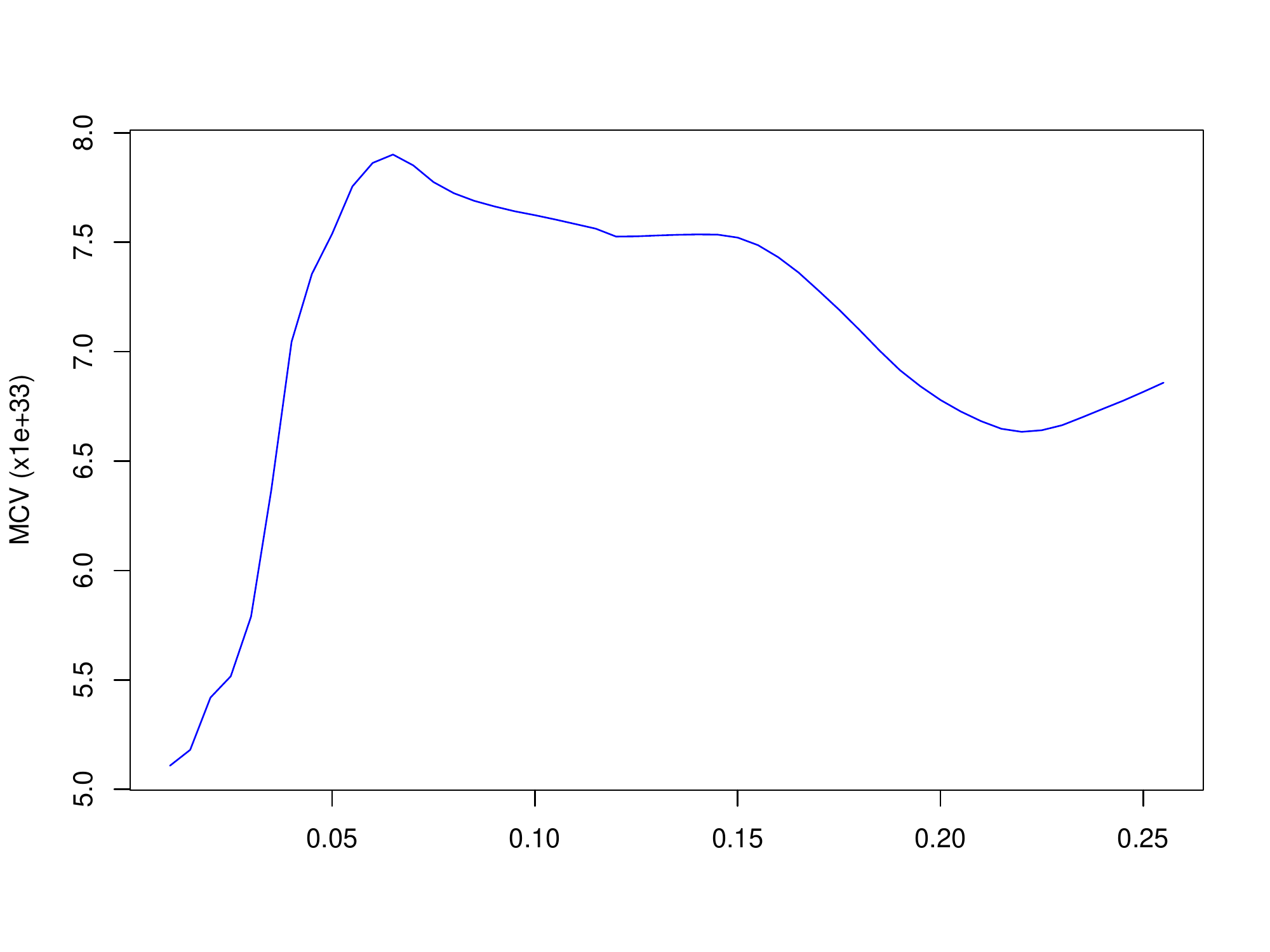}
        }\\
				  \subfigure[Thule]{
            \includegraphics[width=0.45\textwidth]{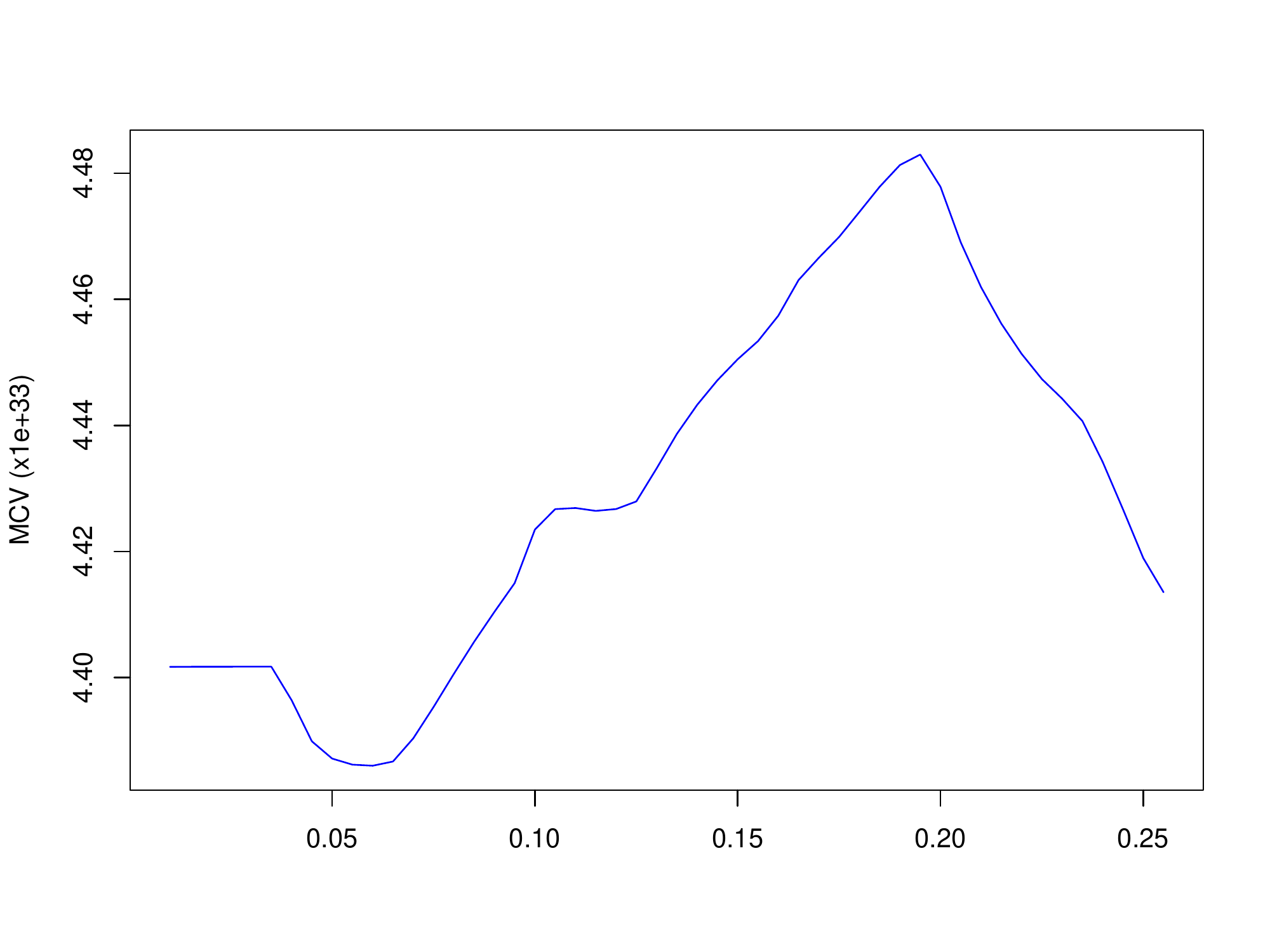}
        }
        \subfigure[Toronto]{
            \includegraphics[width=0.45\textwidth]{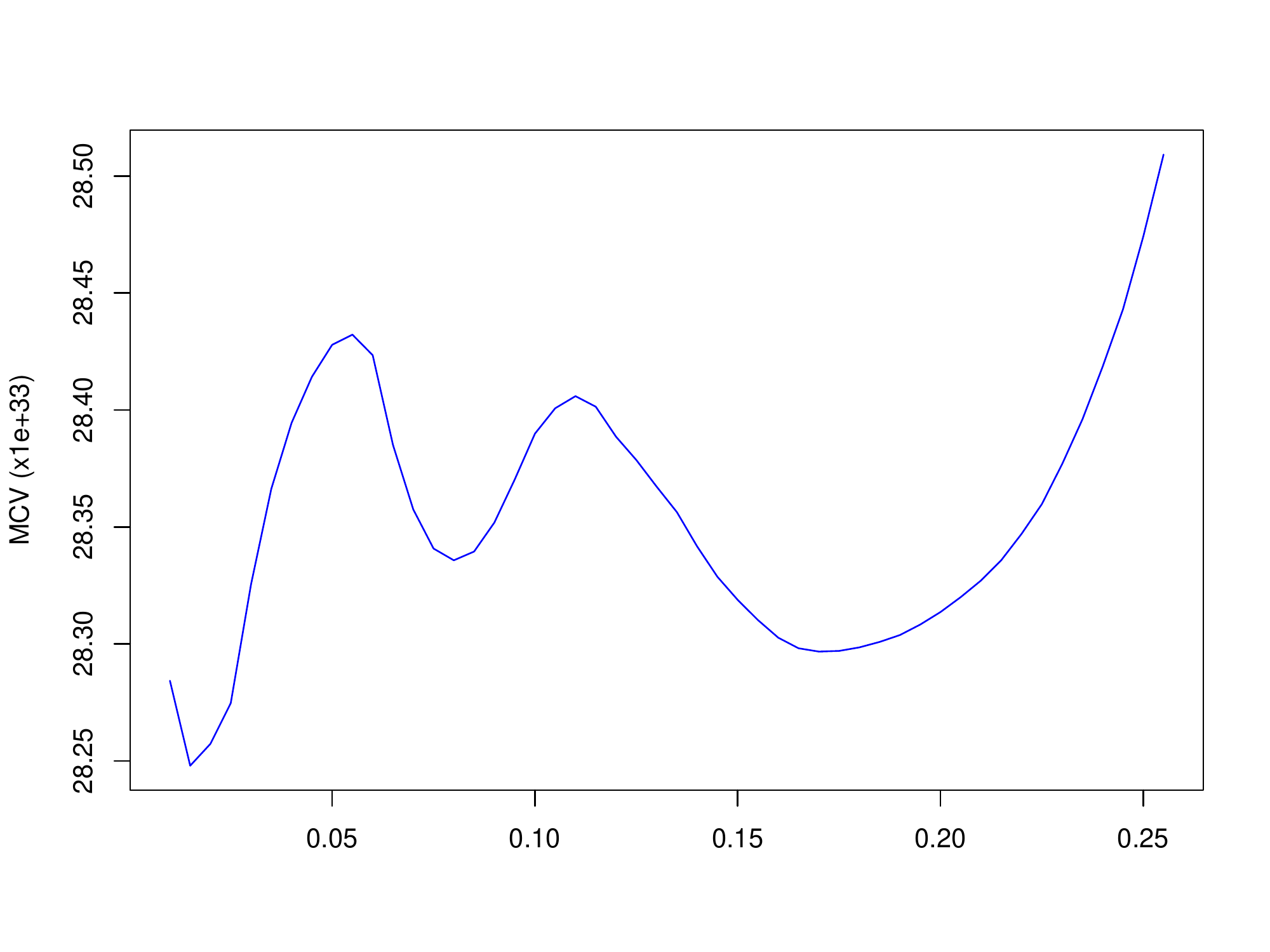}
        }
	\end{center}
	\caption{Modified Cross Validation Criterion for a range of 50 bandwidths (between 0.01 and 0.25 in steps of 0.005). Panel (a) has no minimum. For Panel (b) the first two minima are located at $0.11$ and $0.22$; for Panel (c), they can be found at $0.06$ and $0.12$; for Panel (d) at $0.015$ and $0.085$.}
			\label{fig:bandwidths}
\end{figure}

The MCV procedure has to be applied with care. The range of possible bandwidths over which we minimize the criterion can have a major effect on the resulting optimal bandwidth. The MCV criterion function can have multiple local minima or, in some cases, the function can be monotonically increasing such that it always selects the smallest possible bandwidth. The latter can occur if the values contained in the range of possible bandwidths are too small, but it can also happen using a reasonable grid. To illustrate, in our analysis we allow for values between 0.01 and 0.25 in steps of 0.005. This yields a total of 50 possible bandwidths. We plot the criterion as function of the bandwidth in Figure \ref{fig:bandwidths}. For all series except the Jungfraujoch we can observe at least two local minima which we collect in the caption. The bandwidth choice depends on the context of the study and has to be made by the user. In our case, we prefer a bandwidth that is small enough to allow us to see developments in the trend curve that are missed by the linear trend approach but which produces a reasonably smooth estimate. For Lauder, we therefore select the first bandwidth and for Thule and Toronto the second. In the Jungfraujoch case, the criterion is monotonically increasing. There is a kink at 0.03; the resulting trend estimate with this bandwidth still contains a lot of variation. Since we are interested in longer term movements, we select a slightly larger value of 0.05.

\begin{figure}[h!]
     \begin{center}
        \subfigure[Jungfraujoch]{
            \includegraphics[width=0.47\textwidth,clip, trim = {0.2cm 1cm 0cm 2cm}]{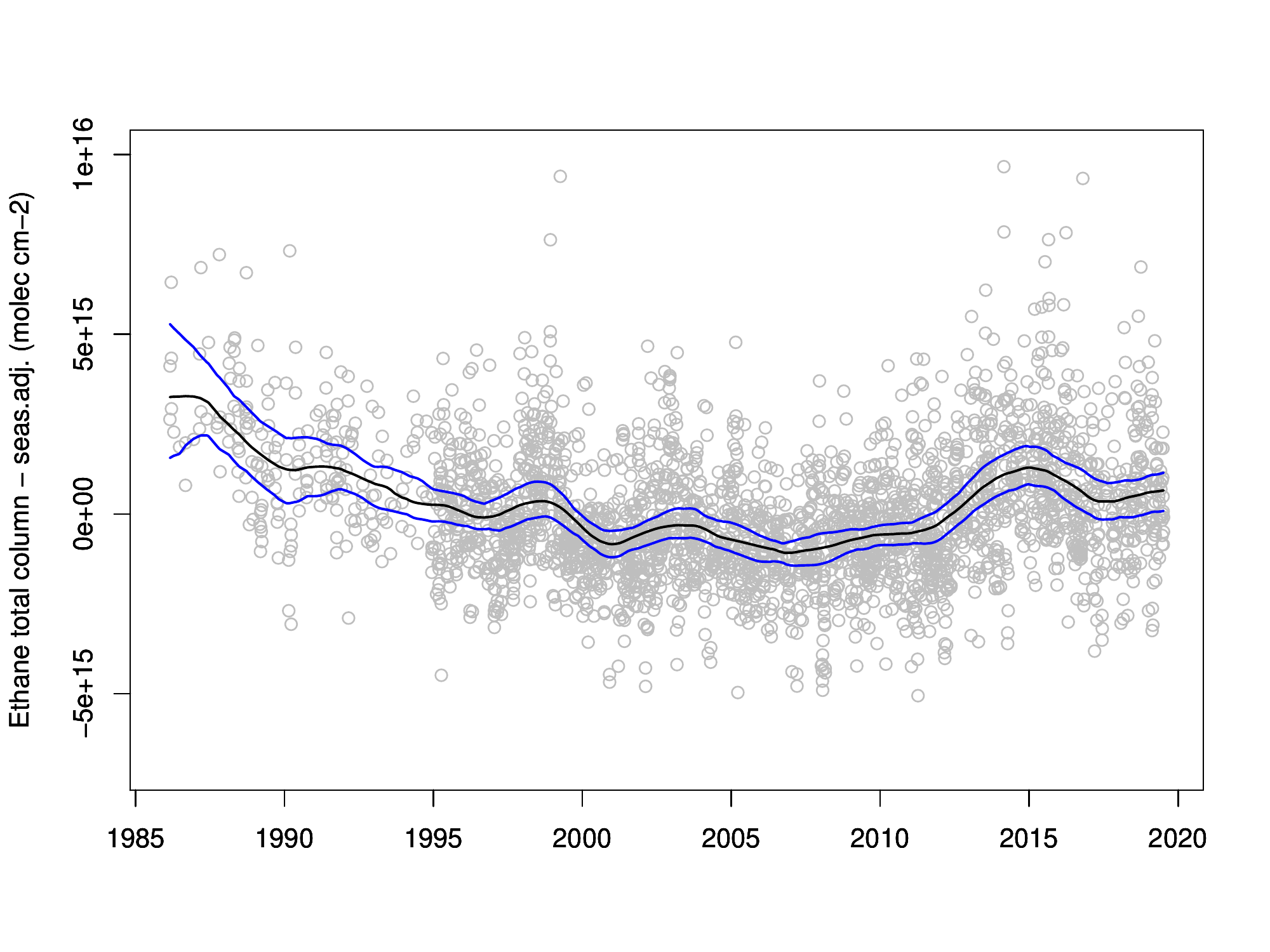}
        }
        \subfigure[Lauder]{
            \includegraphics[width=0.47\textwidth,clip, trim = {0cm 1cm 0cm 2cm}]{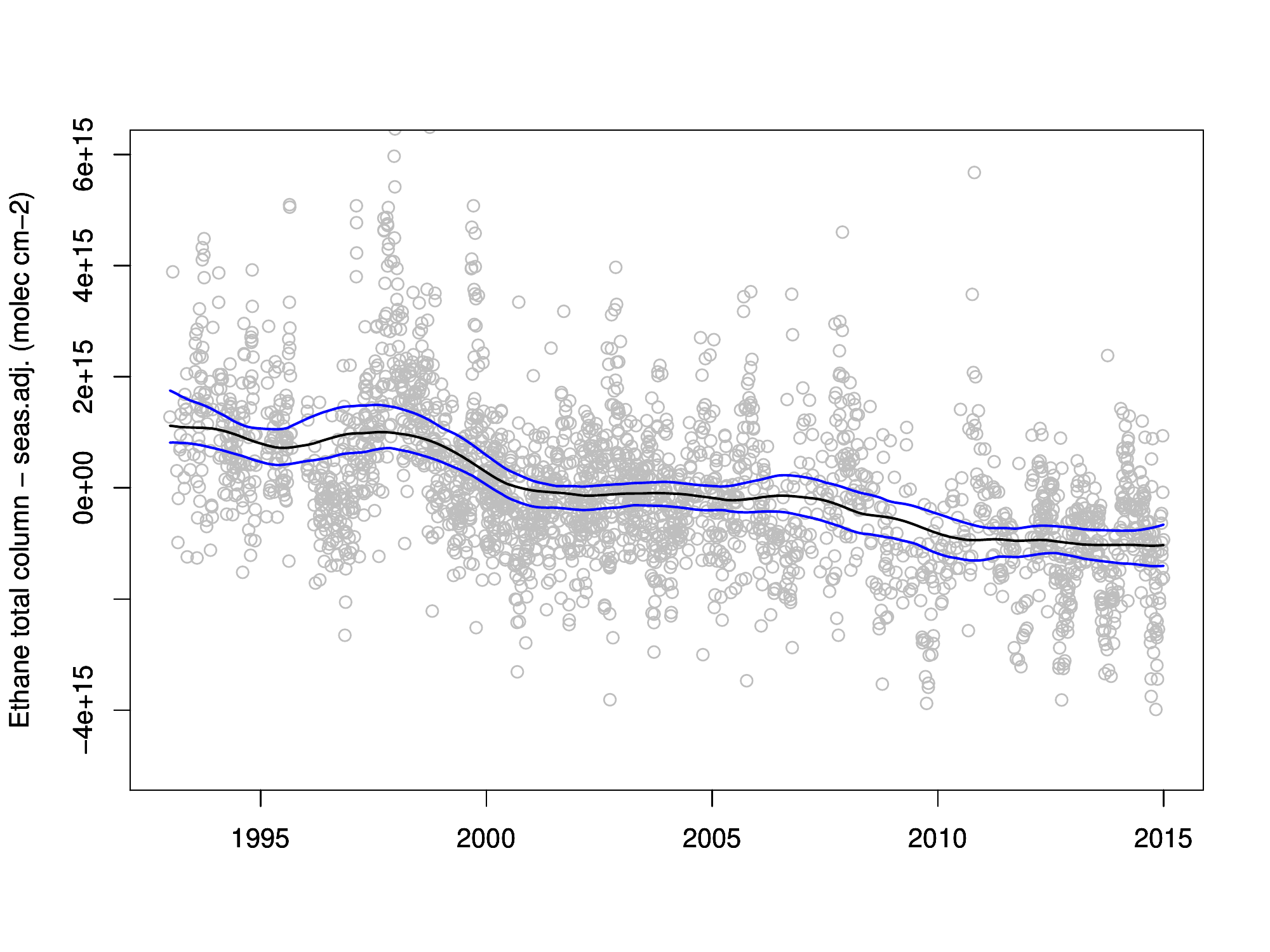}
        }\\
	      \subfigure[Thule]{
            \includegraphics[width=0.47\textwidth,clip, trim = {0.2cm 1cm 0cm 2cm}]{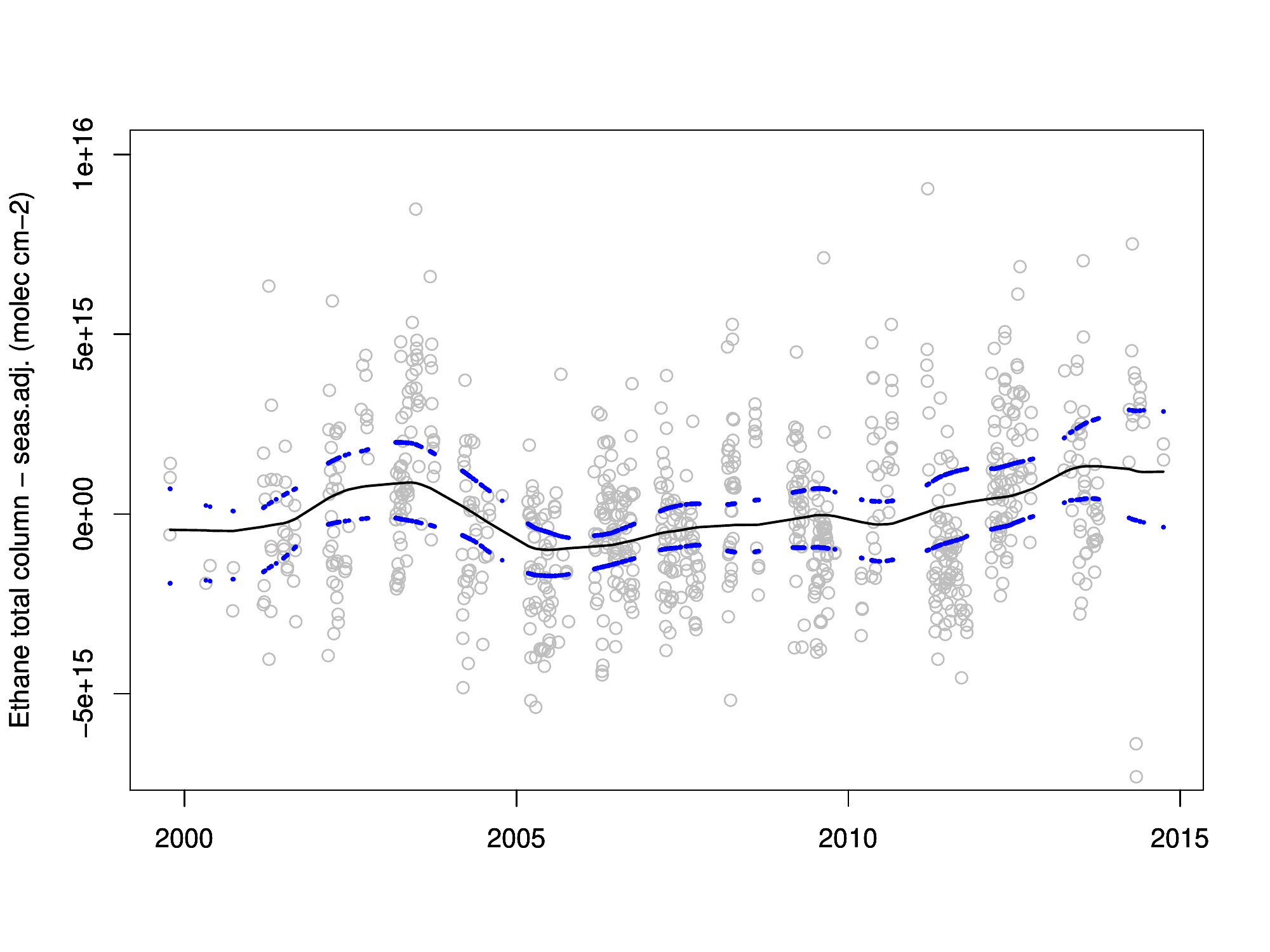}
        }
        \subfigure[Toronto]{
            \includegraphics[width=0.47\textwidth,clip, trim = {0.2cm 1cm 0cm 2cm}]{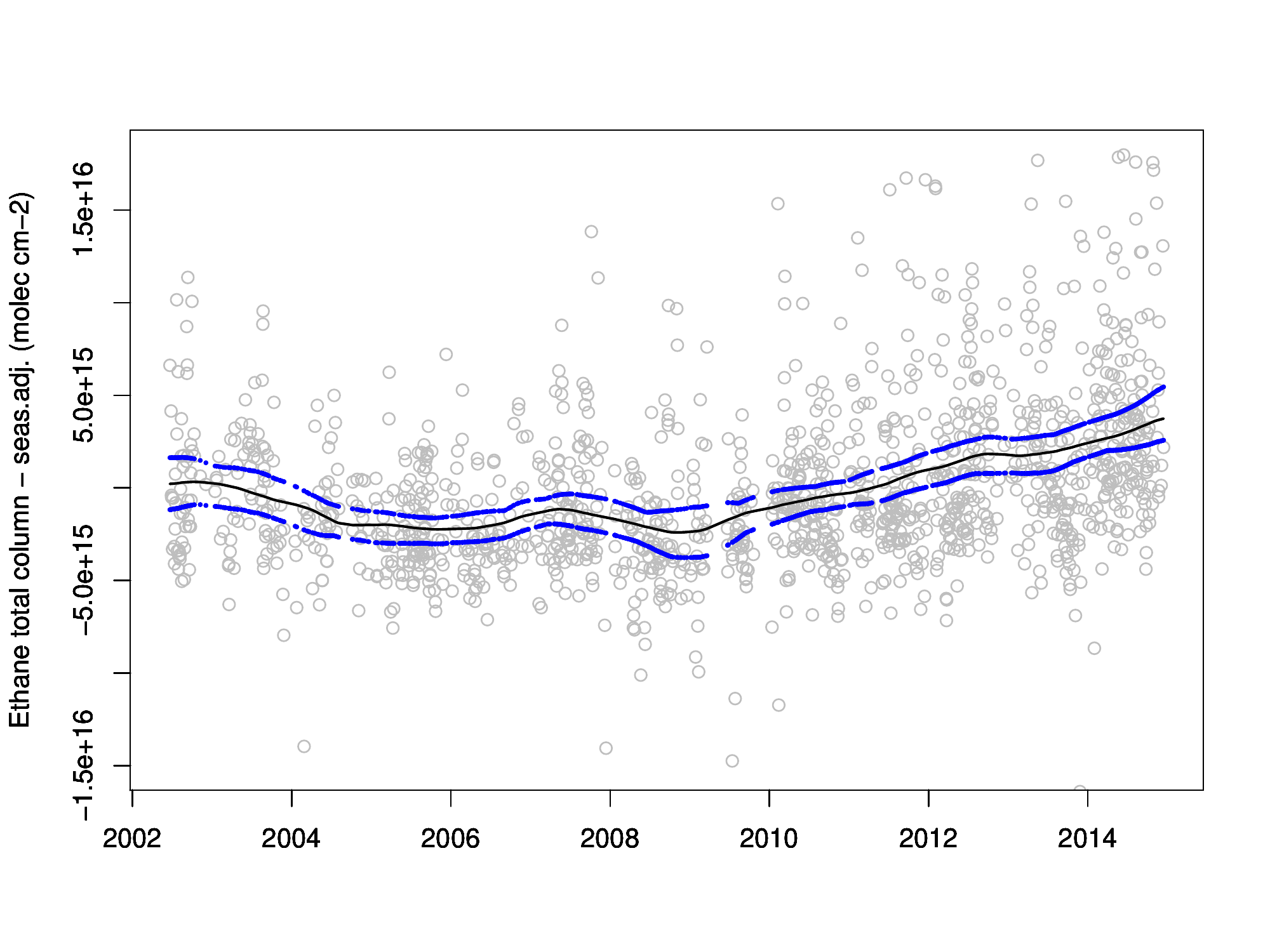}
        }
    \end{center}
    \caption{
        This figure shows the data (gray circles), the nonparametric trend functions (black) and the 95\% simultaneous confidence bands (blue).
     }
   \label{fig:nonparametric}
\end{figure}

Figure \ref{fig:nonparametric} plots the seasonally adjusted data points and the nonparametric trend with the 95\% simultaneous confidence bands in blue. If we follow the movements of the Jungfraujoch trend curve in Panel (a), we see local peaks around the year of 1998 and 2002-2003, which were not visible in the previous analysis. Capturing these two events is possible thanks to the flexibility of the nonparametric approach. A similar peak in 1998 is also visible in the Lauder series (Panel (b)) and in 2002-2003 in the Thule series (Panel (c)). The peaks can be attributed to boreal forest fires which were taking place mainly in Russia during both periods. Geophysical studies have investigated these events in association with anomalies in carbon monoxide emissions \citep{Yurganov1, Yurganov2}. In such fires, carbon monoxide is co-emitted with ethane, such that these events are likely explanations for the peaks we observe.  

In addition, we observe a significant upward trend towards the end of the sample period after a minimum has been reached in 2006 for Jungfraujoch and Thule and around 2009 for Toronto (Panel (d)). This is in line with the parametric analysis and cannot be observed at Lauder. Looking at the confidence bands for the three upward trending series after their minimum has been reached, it is impossible to completely embed a horizontal line into the bands, signaling strong evidence of a nonzero upward trend. A more recent development is the slow down of the upward trend resulting in a peak around 2015. This is a novel finding due to the longer range of our sample. A potential explanation could be the drastic drop in oil prices which occurred in late 2014. Lower oil prices will likely have an impact on the oil and gas industry making it less profitable to exploit shale gas wells.

\subsection{Inference on trend shapes}
\label{sec:trend_shape}
Based on the trend estimates from previous sections, we are interested in particular features of the trend curve. Having in mind the shape of the trends that we discovered, one important feature for our analysis is the local minimum in 2006 of the trend in the Jungfraujoch ethane column series. All other ethane series from the Northern Hemisphere also display a (local) minimum. In the Thule trend estimate, it is located in 2005 and in Toronto in 2008. Therefore, we are interested in the uncertainty around the location of such minima. In order to investigate this issue, we again rely on the autoregressive wild bootstrap method presented above. This is discussed in the first part of this section. The analysis can equally be applied to a local maximum of the trend curve, it is not restricted to the analysis of local minima.

Another interesting feature is the resulting post-minimum upward trend. We have a closer look at the specific form of this trend in the second part. Specifically, we suggest two formal tests; one will compare the nonparametric trend to a linear trend and the other one tests for monotonic behavior in the nonparametric trend. All approaches are applied to investigate the trend in the Jungfraujoch, Thule and Toronto time series. As Lauder does not show the same trend pattern, we drop it for the remainder of the paper.

\subsubsection{Analyzing the locations of extrema}
We are interested in the minimum of the trend estimate, which we denote by $\hat{g}_{min}$, and its location by $t_{min}$. Our goal is to construct a confidence interval for $t_{min}$. For this, we use the autoregressive wild bootstrap to construct bootstrap observations in a similar vein that presented in Algorithm 1. To the bootstrap observations we then apply the nonparametric estimator and determine the location of the local minimum for each bootstrap trend closest to $t_{min}$ - the original minimum - and denote it by $t_{min}^{\ast}$. We give the bootstrap algorithm in Appendix A of the supplementary material.

The proposed analysis can be used to obtain further evidence on the location of a potential trend reversal and the results can be compared to the break location found in the linear trend analysis discussed in Section \ref{sec:parametric}. This new approach is less robust in a sense that it is sensitive to the choice of bandwidth that was used to generate the nonparametric trend estimate. It is, however, much more flexible and less restrictive than the break point detection, as we do not force the trend before and after the minimum to be linear.  

The minimum of the estimated Jungfraujoch trend is located at 2006.86 (10.11.2006). When applying the adapted autoregressive wild bootstrap to obtain 95\% confidence intervals around that location, we find 2006.52 to 2007.38 (08.07.2006 to 16.05.2007), which lies completely within the confidence interval obtained for the break location in Section 3 (2005.66 to 2007.04). It is a good sign that we obtain qualitatively similar result from these two different approaches. The nonparametric approach with this choice of the bandwidth parameter results in a smooth trend, while the parametric specification includes an abrupt break through which the minimum is defined. Similar results are obtained for the Toronto ethane series with a minimum in 2008.84 (30.10.2008) and the 95\% confidence interval ranging from 2007.81 to 2009.53 (23.10.2007 to 12.07.2009), and for Thule with a minimum in 2005.50 (30.06.2005) and confidence intervals ranging from 2005.17 to 2007.40 (02.03.2005 to 23.05.2007).  

\subsubsection{A bootstrap-based specification test} 
When comparing both approaches, the (piecewise) linear and the nonparametric one permitting any smooth nonlinear shape, an obvious question arises as to whether we can say more about the appropriateness of the two trend shapes. While the linear trend has some desirable properties -- e.g. we get an estimate of the average annual decrease or increase in the data -- it might be too restrictive to model the underlying true trend. With the nonparametric approach, we get a better understanding of the true trend shape. Due to its flexibility, however, we do not obtain parameter estimates to measure and compare trends. It can nevertheless be seen as a tool to investigate the plausibility of a linear trend in the different series or subsamples.


\citet{Kapetanios} designs a bootstrap-based test which can be used to test for parameter constancy under the null hypothesis against smoothly occurring structural change. Based on this work, we propose a modification of the test which is able to provide evidence if a certain parametric shape is appropriate to describe the trend in the data at hand. 

We first introduce the test in a general framework. The more specific case of linearity will be discussed later. For the general framework, consider the following null hypothesis:
\begin{equation*}
\textnormal{H}_0:g(t)=g_0\left(\boldsymbol{\theta},t\right)\;\;\;\;\forall t\in\mathcal{G}_m,
\end{equation*}   
where $g_0(\boldsymbol{\theta},\cdot)$ belongs to a parametric family $\mathbf{G}=\{g(\boldsymbol{\theta},\cdot); \boldsymbol{\theta} \in \Theta \subset \mathbb{R}^d\}$ with $d$ being the number of parameters in $\boldsymbol{\theta}$. Further, the set $\mathcal{G}_m=\left\{t_1,t_2,...,t_m\right\}$ contains the time points for which the hypothesis should be tested. Under the alternative, the trend does not follow the parametric shape given by $g_0(\tau)$, but can be expressed as in model \eqref{eq:model_nonpara}. As a special case of the test, which is of particular interest in our application, we can consider the linear trend function $g_0(t)=\alpha+\beta t$ such that $\boldsymbol{\theta}=\left(\alpha,\beta\right)$ and $d=2$. A similar framework is considered in \citet{Wang}, \citet{Zhang} as well as in \citet{Lyubchich}. The proposed tests are, however, designed for equally spaced observations and, therefore, not easily applicable to series with missing data. 

We use an adaption of the test statistic in \citet{Kapetanios}
\begin{equation}
Q_{t}=\left(\hat{g}(t/n)-g_0(\widehat{\boldsymbol{\theta}},t)\right)^2,
\label{eq:test_stat}
\end{equation}
where $\hat{g}(t/n)$ denotes the nonparametric kernel estimator, as before, and $\widehat{\boldsymbol{\theta}}$ denotes the parameter estimates under the null hypothesis. The type of estimator we choose under the null hypothesis depends on the specific case and the form of the parametric function. In the linear trend case, we can use OLS to obtain estimates $\hat{\alpha}$ and $\hat{\beta}$. As the subscript $t$ shows, this test statistic is pointwise. Since we are interested in the trend over time, we follow \citet{Kapetanios} and use the two summary statistics for the set $\mathcal{G}_m=\left\{t_1,t_2,...,t_m\right\}$
\begin{align}
Q_{ave}&=\frac{1}{m}\sum_{j=1}^m Q_{t_j}\\
Q_{sup}&=\sup_j Q_{t_j}.
\end{align}

To obtain critical values for the test statistics, we rely again on the autoregressive wild bootstrap method.\footnote{Next to bootstrapping, \citet{Kapetanios} also investigates asymptotic tests which show a particularly poor performance. We can therefore expect that a bootstrap-based test is also preferred in our slightly different set-up.} 

Our test can loosely be interpreted as a functional extension of a traditional specification test in the spirit of \citet{Hausman}, where we have one estimator that is consistent under both the null and alternative hypothesis - here the nonparametric one - and one that is more efficient under the null hypothesis - the correctly specified parametric model - but inconsistent under the alternative. Therefore, we expect the two estimators to be close to each other under the null, for all considered $t$, and hence $Q_t$ will be close to zero. Under the alternative, the estimators will differ, leading $Q_t$ to diverge at some $t$. We then detect these divergences by aggregating the $Q_t$ statistics over all $t \in \mathcal{G}_m$ in one of the two ways described above.

The exact specification of the set $\mathcal{G}_m$ depends on the application at hand. In practice, often a set of several consecutive points is needed to be able to estimate the parameters under the null hypothesis. This is the case, for example, with the linear trend application that we focus on in the remainder of the section.  

We choose the set $\mathcal{G}_m$ in such a way that it covers the period of increase in the ethane trends. Specifically, we select the starting point of $G_m$ as the minimum of the nonparametrically estimated trend, which we have determined in the previous section. The end point coincides with the end of the sample. While one could clearly take any starting point, testing for linearity appears counter-intuitive if we start before the minimum. This would be equivalent to asking whether a line with a kink could be described as linear. Since this null hypothesis would surely be rejected, we consider a more interesting part of the sample.\footnote{Alternatively, one could test whether a piecewise linear model is appropriate for a larger part of the sample, if at least one is comfortable with using the nonparametric estimator for modeling an abrupt change by a smooth approximation. The mechanics of such a test are the same as the test we consider here.} We connect the nonparametrically estimated part of the trend to the linear part imposed by the null hypothesis at the point $\hat{g}_{min}$, thereby testing whether the trend after this point is linear. To connect the two subsamples, the intercept is determined by the value of the trend before $\hat{g}_{min}$, and only the slope parameter is estimated by OLS. For the calculation of the test statistic \eqref{eq:test_stat} we use as $g_0(\widehat{\boldsymbol{\theta}},t)$ the best fitting linear trend line that goes through the minimum for all $t$ in $\mathcal{G}_m$.

The results are summarized in Panel A of Table \ref{tab:lin_mono}. We report the values of the two different versions of the test as well as the bootstrap critical values and the resulting $p$-values. The test leads to a rejection of the null hypothesis of a linear trend for the Jungfraujoch series at a 1\% significance level, while it does not reject linearity for Thule and Toronto at any reasonable level. 

\begin{table}[t]
\centering
\begin{tabular}{lccccccc}
\noalign{\smallskip} \hline \hline
\multicolumn{8}{c}{\textbf{A: Linearity test}} \\
\noalign{\smallskip} \hline
 & $Q_{ave}$ & \textit{CV}$_{ave}$& $p_{ave}$ & $Q_{sup}$ & \textit{CV}$_{sup}$&$p_{sup}$ & \\
\hline
\noalign{\smallskip}
Jungfraujoch  & 4.616$\times10^{28}$ & 3.434$\times10^{28}$  & 0.014 & 3.803$\times10^{29}$   & 1.637$\times10^{29}$& 0.000 &  \\
Thule    & 3.564$\times10^{28}$  &  1.124$\times10^{29}$ & 0.563 & 2.394$\times10^{29}$ & 7.105$\times10^{29}$ & 0.379 &  \\
Toronto  &   1.659$\times10^{29}$ & 2.917$\times10^{29}$ & 0.200 &  5.507$\times10^{29}$  & 1.378$\times10^{30}$ &  0.438  &       \\
\noalign{\smallskip} \hline 
\multicolumn{8}{c}{\textbf{B: Monotonicity test}} \\
\noalign{\smallskip} \hline
 & $U_{1}$ & \textit{CV}$_{1}$& $p_{1}$ & $U_{2}$ & \textit{CV}$_{2}$&$p_{2}$ & $h_{U}$ \\
\hline
\noalign{\smallskip}
Jungfraujoch  & 0.177 & 0.131  & 0.002 & 5.247$\times10^{14}$   & 4.084$\times10^{14}$& 0.010 & 0.101 \\
Thule    & 0.140  &  0.243 & 0.453 & 2.268$\times10^{14}$ & 9.688$\times10^{14}$ & 0.939 & 0.131 \\
Toronto  &   0.003 & 0.152 & 1.000 &  1.423$\times10^{13}$  &1.234$\times10^{15}$ &  1.000  & 0.117       \\
\noalign{\smallskip} \hline \noalign{\smallskip}
\end{tabular}
\caption{Inference on trend shapes. (A) Results of the linearity test statistics $Q_{ave}$ and $Q_{sup}$ as well as the corresponding critical values (\textit{CV}) and $p$-values. (B) Results of the two monotonicitiy test statistics $U_{1}$ and $U_{2}$ as well as the corresponding critical values (\textit{CV}) and $p$-values. In the last column, the bandwidths of the tests are reported.}
\label{tab:lin_mono}
\end{table}

\subsubsection{Two tests for monotonicity}
In the previous section we proposed a bootstrap-based test to investigate if the trend can be best described by a specific parametric shape - in this case linearity - or by the unrestricted nonparametric alternative. In some applications, however, the question whether the trend has been monotonically increasing or decreasing over a certain period can already be enough evidence. In the case of the ethane series, we are mainly interested in establishing an upward trend in the post-minimum period of the sample. Therefore, we propose to additionally use two tests for monotonicity.

The test considers a monotonically increasing trend function under the null hypothesis. The alternative is the same as before, a nonparametric unrestricted trend. Formally, this can be written as:
\begin{equation*}
\textnormal{H}_0:g(\cdot)\; \textnormal{is an increasing function on}\; \mathcal{I},
\end{equation*}
or, since under the given smoothness assumptions the function $g(\cdot)$ is differentiable:
\begin{equation*}
\textnormal{H}_0:g'(t/T)\geq 0\;\;\;\;\forall t\in\mathcal{I}.
\end{equation*}
In this case, the set $\mathcal{I}$ must be a compact interval in the domain of the function $g(\cdot)$. The paper by \citet{GSV} proposes the following test statistic to test the above null hypothesis, for $t\in\mathcal{I}$:
\begin{equation}
U_{1,t}=-\frac{2}{T(T-1)}\sum_{1\leq i<j\leq T}\operatorname{sign}\left(y_j-y_i\right)\frac{1}{h_U}K\left(\frac{i/n-t/n}{h_U}\right)\frac{1}{h_U}K\left(\frac{j/n-t/n}{h_U}\right)
\label{eq:U_stat}
\end{equation} 
with
\begin{equation*}
\operatorname{sign}(x) = \left\{
\begin{array}{ll}
1  & \text{if $x>0$} \\
0  & \text{if $x=0$} \\
-1 & \text{if $x<0$} 
\end{array} \right.
\end{equation*}
As kernel function, we use $K(x)=0.75\left(1-x^2\right)$ for $-1<x<1$ and $0$ otherwise, as \citet{GSV} suggests. We also follow their bandwidth recommendation $h_U=0.5T^{-1/5}$. The test is based on the idea that for an increasing function, increments will be positive and thus, the test statistic should satisfy $U_{1,t}\leq 0$ for most $t\in\mathcal{I}$ under the null. This can be easily verified as $U_t^t$ sums over weighted differences of observations $\left(y_j-y_i\right)$ such that $j>i$; or more precisely, it sums over the sign thereof. The test statistic $U_{1,t}$ corresponds to one point in the interval of interest, $\mathcal{I}$, similar to the test statistic $Q_t$ in \eqref{eq:test_stat}. As summary statistic, \citet{GSV} propose a supremum statistic
\begin{equation}
U_{1}=\sup_{t\in\mathcal{I}}U_{1,t}.
\end{equation}

Additionally, we use a second test to support our findings. This second test is proposed in \citet{Chetverikov}. The difference compared to \eqref{eq:U_stat} is the use of the sign function, which is omitted in this version of the test. The full differences and not only their sign will be accounted for. This gives the following test statistic:
\begin{equation}
U_{2,t}=-\frac{2}{T(T-1)}\sum_{1\leq i<j\leq T}\left(y_j-y_i\right)\frac{1}{h_U}K\left(\frac{i/n-t/n}{h_U}\right)\frac{1}{h_U}K\left(\frac{j/n-t/n}{h_U}\right),
\label{eq:U_stat2}
\end{equation} 
which we apply with the same specifications as we use for $U_{1,t}$. Again, this statistic is negative under the null hypothesis due to the same reason as above. In line with the above procedure, we calculate summary test statistics $U_{2}$ whose exact definition follow in analogy to $U_{1}$. 

To obtain critical values, we rely once again on the autoregressive wild bootstrap. In this case, we need to make one adjustment to the bootstrap algorithm for the nonparametric trend reported in Appendix A, which is that the trend is set to zero in the construction of bootstrap observations. This makes sure that the null hypothesis is satisfied. 

Coming back to the original research question and motivation for this test, we now investigate the post-minimum nonparametric trend of the three ethane series obtained at Jungfraujoch, Thule and Toronto. After having rejected linearity for the Jungfraujoch location, this test helps us to establish whether there has been a monotonic upward trend in the series since their respective minimum. Thus, the set $\mathcal{I}$ over which we test for monotonicity coincides with the set $G_m$ we selected for the linearity test above. 

The results are summarized in Panel B of Table \ref{tab:lin_mono}. We report the values of the two different versions of the test as well as the bootstrap critical values and the resulting $p$-values. The test provides evidence that the Jungfraujoch post-minimum trend is not monotonically increasing. This result is likely driven by the slow down in the estimated trend around 2015. For the other two locations, we cannot reject the null hypothesis and conclude that the post-minimum trend in the ethane burden at Thule and Toronto is monotonically increasing, which is in line with the results from the (post-minimum) linearity test.

\section{Conclusion}
\label{sec:conclusion}
We analyze trends and trend reversals in a set of four time series of ethane total columns. Three series are obtained from measurement stations located in the Northern Hemisphere: Jungfraujoch in the Swiss Alps, Thule in Greenland, and Toronto in Canada. One series is taken in Lauder which is located in the Southern Hemisphere. The stations record daily observations of ethane abundance in the atmosphere. Depending on the conditions, however, measurements cannot be made during cloudy days resulting in time series with data available on about one day in three. This is a limitation frequently encountered in (climatological) time series, which causes problems when constructing confidence intervals around the trend estimate. 

This paper proposes two approaches for trend analysis in such settings. First, a broken linear trend model is estimated with unknown break date. Piecewise linear trends have the advantage of being easy to estimate and interpret. However, imposing linearity may obscure important features that are not well captured by linearity. Second, we move to a nonlinear and nonparametric model. This model allows us to capture much richer features at the expense of more complicated estimation and interpretation. For the construction of confidence intervals in the nonparametric model, we use an autoregressive wild bootstrap method. Additionally, we propose several diagnostic tools to investigate the shape of the resulting trend.

There is a significant upward trend in atmospheric ethane, starting around 2006/2007. This finding is confirmed by both approaches as the break of the linear model and the local minimum of the nonparametric approach is in located in 2006. The subsequent results of a formal test for linearity indicate that a linear trend is not appropriate for the post-minimum period of the Jungfraujoch and Thule series. In addition, the nonparametric estimation reveals trend functions which exhibit local maxima around the years of 1998 and 2002-2003 which coincide with boreal forest fires in Russia which were not captured by the linear model.

The two approaches proposed in this paper should be viewed as complimentary rather than competing methods. The simplicity of the broken linear trend model allows us to indicate a numerical value for the slope parameter, summarizing the development of the trend over a particular period. Even if one does not truly believe in linearity of the trend, it may still prove to be a useful approximation given its simplicity. Alternatively, the complexity of the nonlinear approach has the potential of providing us with additional information and capturing features obscured by the linear model. At the same time it can be used to confirm previous findings and to judge the plausibility and appropriateness of the linear trend model.

A limitation of the piecewise linear trend model presented here is that it can accommodate only one break, putting it at a natural disadvantage to the more flexible nonparametric approach. Indeed, estimation of broken linear trend models with multiple breaks at unknown locations can be estimated using, for instance, the methods proposed in \citet{BaiPerron}, which also allow one to test for the number of breaks in the trend. However, constructing confidence intervals for the locations of multiple breaks is more complicated in such models, and the bootstrap method for a single break is not easily adapted. The extension of the bootstrap methodology to multiple breaks is left for future research.

\FloatBarrier
\newpage

\begin{appendices}
\section{Technical appendix}
\label{sec:AppendixB}
\subsection{Linear trend estimation}
The first step in the procedure is to estimate the break date. Subsequently, given a candidate break date $T_c$, estimates of $\left(\alpha,\beta,\delta,s_t\right)$ are obtained by minimizing the following sum of squared residuals
\begin{equation}
\left(\hat{\alpha}_{T_c},\hat{\beta}_{T_c},\hat{\delta}_{T_c},\hat{s}_{t,T_c}\right)=\underset{\alpha,\beta,\delta,s_t}{\operatorname{argmin}}\sum_{t=1}^{T}M_t\left(y_t-\alpha-\beta t-\delta D_{t,T_c}-s_{t}\right)^2,
\end{equation}
where our notations with subscripts $T_c$ makes explicit that these estimates are for a candidate break date $T_c$, and not yet the final parameter estimates. The two step procedure is as follows. In the first step, the estimation of the break location, we construct a sum of squared residuals for every admissible break date candidate $T_c$. The minimum over all possible candidates gives us the estimated break date. Hence, for $\Lambda$ denoting the set of all possible break locations, we have
\begin{equation}
\hat{T}_1=\underset{T_c\in \Lambda}{\operatorname{argmin}}\sum_{t=1}^{T}M_t\left(y_t-\hat{\alpha}_{T_c}-\hat{\beta}_{T_c} t-\hat{\delta}_{T_c} D_{t,T_c}-\hat{s}_{t,T_c}\right)^2,
\label{eq:breakdate}
\end{equation}
where $(\hat{\alpha}_{T_c},\hat{\beta}_{T_c},\hat{\delta}_{T_c},\hat{s}_{t,T_c})$ are determined as described above. Once we have obtained $\hat{T}_1$, we construct the corresponding least-squares parameter estimates in the second step. To be consistent with the notation, these will be denoted as $(\hat{\alpha},\hat{\beta},\hat{\delta},\hat{s}_t)=(\hat{\alpha}_{\hat{T}_1},\hat{\beta}_{\hat{T}_1},\hat{\delta}_{\hat{T}_1},\hat{s}_{t,\hat{T}_1})$. In the next sections, we will show how to use a bootstrap approach to construct confidence intervals fr both the parameter estimates as well as the break location.

\subsection{Confidence intervals for the linear trend model}
Given the parameter estimates and the estimated break location, we need some measure of uncertainty to assess the significance of our findings. A major difficulty with climate time series is the presence of serial correlation. An additional complication arises because these time series often have observations that are unequally spaced over the sample period $t=1,...,T$. To overcome these difficulties, we propose a bootstrap method which is well-established in the econometrics and statistics literature and provides accurate confidence intervals even in small samples. This bootstrap approach works in the presence of serial correlation and it can be applied to unequally spaced data. In addition, it remains valid when there are possible changes in variance of the residuals.

To form bootstrap samples, the standard bootstrap method - the i.i.d. bootstrap - draws randomly and with replacement from the residuals and, thereby, destroys both the dependence structure and possible time variations in the variance. Such bootstrap sample will not mimic the original series of residuals and the general principle, on which bootstrap methods are based, is violated. Instead, we should construct bootstrap errors which have the same pattern of correlation, variance changes and missing data as the original set of residuals. The autoregressive wild bootstrap, which is proposed in the context of nonparametric trend estimation in \citet{FSU}, is to our knowledge the best way to achieve this goal. In particular, in the presence of serial correlation compared to its competitors - sieve or block bootstrap methods - it holds a clear advantage as it has a natural way of handling missing data. No adjustments are needed for it to reproduce the missing data pattern in the original sample.

In the remainder of this section, we provide the algorithm to construct bootstrap confidence intervals for the break date estimate. Explanations on how to adapt this approach to form confidence intervals for the parameter estimates of slope and intercept follow afterwards.
\setcounter{algorithm}{1}
\begin{algorithm}[Autoregressive Wild Bootstrap - Break location]
$\phantom{2}$
\begin{enumerate}
\item Calculate residuals from the estimation of model \eqref{eq:genmodel} with the trend $d_t$ specified by \eqref{model}. Impose a break at $\hat{T}_1$. For $t=1,...,T$,  \begin{equation*}\hat{u}_t=M_t\left(y_t-\hat{\alpha}-\hat{\beta}t-\hat{\delta}D_{t,\hat{T}_1}-\hat{s}_t\right).\end{equation*} 
\item For $0 < \gamma < 1$, generate $\nu_1^\ast,\ldots,\nu_n^\ast$ as i.i.d.~$\mathcal{N}(0,1-\gamma^2)$ and let $\xi_t^* = \gamma \xi_{t-1}^* + \nu_t^*$ for $t=2,\ldots,T$. Take $\xi_1^* \sim \mathcal{N}(0,1)$ to ensure stationarity of $\{\xi_t^*\}$.
\item Calculate the bootstrap errors $u_t^{\ast}=M_t\xi_t^{\ast}\hat{u}_t$ and generate the bootstrap sample as \begin{equation*}y_t^\ast=M_t\left(\hat{\alpha}+\hat{\beta}t+\hat{\delta}D_{t,\hat{T}_1}+\hat{s}_t+u_t^\ast\right)\end{equation*} for $t=1,...,T$, using the same estimated coefficients as in Step 1.
\item Determine $\hat{T}_1^{\ast}$ from $y_t^{\ast}$ as in \eqref{eq:breakdate} and store the estimate.
\item Repeat Steps 2 to 4 a total of $B$ times and let 
\begin{equation*}
\hat{q}_{T_1,\alpha}=\inf\left\{u;P^{\ast}\left[\hat{T}^{\ast}_1-\hat{T}_1\leq u\right]\geq  \alpha\right\}
\end{equation*}
denote the $\alpha$-quantile of the $B$ centered bootstrap quantities $(\hat{T}_1^{\ast}-\hat{T}_1)$. 
\end{enumerate}
\end{algorithm}
The confidence intervals are then determined by the $\alpha/2$ and $(1-\alpha/2)$ quantiles as
\begin{equation}
I_{T_1,\alpha}=\left[\hat{T}_1-\hat{q}_{T_1,1-\alpha/2},\hat{T}_1-\hat{q}_{T_1,\alpha/2}\right].
\label{eq:CIintervalsbootstrap}
\end{equation}
The construction of confidence intervals for the parameter estimates requires an adjustment to Step 4. That is, given the estimated break location $\hat{T}_1$, we estimate model \eqref{model} from $y_t^{\ast}$. Bootstrap estimates $\hat{\alpha}^{\ast}$, $\hat{\beta}^{\ast}$ and $\hat{\delta}^{\ast}$ are stored and the corresponding bootstrap distributions and quantiles are constructed as before. The resulting confidence intervals are similar to \eqref{eq:CIintervalsbootstrap}.

The bootstrap algorithm above looks similar to Algorithm 1 from the main text. There is however one important difference. In Algorithm 1 the bootstrap series are constructed under the null hypothesis of no break and a linear trend model \emph{without} break is estimated from the data in Step 1. In the current situation we are assuming that there is a structural break and the model used in Step 1 should reflect this.

\subsection{Nonparametric estimation}
The main goal is to estimate the trend function $g(\cdot)$ and to determine the uncertainty around this estimate. To achieve this goal we propose using a two-stage estimation procedure. In the first stage, $y_t$ is regressed on the Fourier terms and in the second stage, the residuals from the first stage are used to estimate the trend nonparametrically. Denote by $\hat{\epsilon}_t$ the residuals from a regression of $y_t$ on the Fourier terms so that $\hat{\epsilon}_t=M_t\left(y_t-\hat{s}_t\right)$. To estimate the trend function from $\left\{\hat{\epsilon}_t\right\}$, we apply a nonparametric kernel estimator: the local constant Nadaraya-Watson estimator. It is found by minimizing a weighted sum of squares with respect to $g(\cdot)$. This is done for every point $t=1,...,T$. As is standard with this approach, we map the points into the interval $\left(0,1\right)$. Thus, for $\tau\in\left(0,1\right)$, we obtain:
\begin{equation}
\begin{split}
\hat{g}(\tau) &= \argmin_{g(\tau)} \sum_{t=1}^T K\left(\frac{t/T-\tau}{h}\right) M_t \left\{\hat{\epsilon}_t-g(\tau)\right\}^2 \\
&=\left[\sum_{t=1}^T K\left(\frac{t/T-\tau}{h}\right) M_t \right]^{-1} \sum_{t=1}^T K\left(\frac{t/T-\tau}{h}\right)M_t\hat{\epsilon}_t,
\label{eq:lc_estimator}
\end{split}
\end{equation}
where $K(\cdot)$ is a kernel function and $h>0$ is the bandwidth. We propose to use the Epanechnikov kernel which is given by the function $K(x)=\frac{3}{4}(1-x^2)\mathbbm{1}_{\left\{|x|\leq 1\right\}}$. The parameter $h$ is the bandwidth. Unfortunately, data-driven methods for bandwidth selection frequently applied in practice show problems when applied to time series data. Therefore, we propose using a time series version of such a criterion, called modified cross-validation (MCV). It is an adapted version of the original criterion proposed in \citet{ChuMarron}:
\begin{equation}
CV_k(h)=\frac{1}{T}\sum_{t=1}^TM_t\left(\hat{g}_{k,h}\left(\frac{t}{T}\right)-\hat{\epsilon}_t\right)^2,
\label{eq:CV}
\end{equation}
where
\begin{equation}
\hat{g}_{k,h}(t/T)=\frac{(T-2k-1)^{-1}\sum_{t:|t-\tau T|>k}K\left(\frac{t/T-\tau}{h}\right)M_t\hat{\epsilon}_t}{(T-2k-1)^{-1}\sum_{t:|t-\tau T|>k}K\left(\frac{t/T-\tau}{h}\right)M_t},
\end{equation}
is a leave-$(2k+1)$-out version of the leave-one-out estimator, which leaves out the observation receiving the highest weight. The criterion is based on minimizing the sum of squared residuals from a model fit using a candidate bandwidth. To avoid overfitting, the observation receiving the highest weight is left out of the estimation. In this adapted version of \citet{ChuMarron}, $k$ observations around it are left out as well. The optimal bandwidth is found by minimizing this criterion with respect to $h$. We refer to Section \ref{sec:app_nonpara} for the practical illustration.

\subsection{Confidence intervals around the nonparametric trend}
Confidence intervals are needed to say more about the significance of the trend. In the nonparametric setting, the object of interest is the trend function as a whole because there are no parameters that fully describe the trend function - as opposed to slope coefficient in the previous approach. We therefore need a tool to judge the significance over more than one time point, preferably, over the whole sample. Such confidence intervals could be used, for example, to assess whether there has been a significant non-zero upward or downward trend. 

Confidence bands, where the coverage simultaneously holds over more than one point or even the whole sample, can be constructed from pointwise confidence intervals. We therefore first propose a method to construct pointwise confidence intervals for the nonparametric trend estimate. Second, based on these intervals, we suggest a three-step algorithm to transform pointwise intervals into simultaneous confidence bands. For this type of model and the estimators we use, it has been shown in the statistical literature that bootstrap methods are a reliable tool to conduct inference (see e.g. \citet{Buhlmann}, \citet{NP}). Given the difficulties we face with atmospheric time series with irregular sampling over time, we again propose to use the autoregressive wild bootstrap. A minor adjustment compared to the above algorithm has to be made. For completeness, we again specify the full bootstrap algorithm.

\begin{algorithm}[Autoregressive Wild Bootstrap - Nonparametric trend]
$\phantom{3}$
\begin{enumerate}
\item Let $\tilde{g}(\cdot)$ be defined as in \eqref{eq:lc_estimator}, but using bandwidth $\tilde{h}$. Obtain residuals \begin{equation*}
\hat{u}_t = M_t\left(\hat{\epsilon}_t-\tilde{g}(t/T)\right),\;\;\;\;\;t=1,\ldots,T,
\end{equation*}

\item For $0 < \gamma < 1$, generate $\nu_1^\ast,\ldots,\nu_n^\ast$ as i.i.d.~$\mathcal{N}(0,1-\gamma^2)$ and let $\xi_t^* = \gamma \xi_{t-1}^* + \nu_t^*$ for $t=2,\ldots,T$. Take $\xi_1^* \sim \mathcal{N}(0,1)$ to ensure stationarity of $\{\xi_t^*\}$.

\item Calculate the bootstrap errors $u^{\ast}_t$ as $u_t^\ast=M_t\xi_t^\ast\hat{u}_t$ and generate the bootstrap observations by
\begin{equation*}
\hat{\epsilon}^{\ast}_t = M_t\left(\tilde{g}(t/T) + u^{\ast}_t\right),\;\;\;\;\;t=1,...,n,
\end{equation*}
where $\tilde{g}(t/T)$ is the same estimate as in the first step.

\item Obtain the bootstrap estimator $\hat{g}^*(\cdot)$ as defined in \eqref{eq:lc_estimator} using the bootstrap series $\hat{\epsilon}_t^{\ast}$, with the same bandwidth $h$ as used for the original estimate $\hat{g}(\cdot)$.

\item Repeat Steps 2 to 4 a total of $B$ times and let 
\begin{equation*}
\hat{q}_{\alpha}(\tau)=\inf\left\{u;P^{\ast}\left[\hat{g}^{\ast}(\tau)-\tilde{g}(\tau)\leq u\right]\geq  \alpha\right\}
\end{equation*}
denote the $\alpha$-quantile of the $B$ centered bootstrap statistics $\hat{g}^{\ast}(\tau)-\tilde{g}(\tau)$. These bootstrap quantiles are used to construct confidence bands as described below.
\end{enumerate}  
\end{algorithm}
Note that in Step 1 of the above algorithm, a different bandwidth is used to perform the nonparametric estimation. We suggest to use the larger bandwidth $\tilde{h}=0.5h^{5/9}$. This produces an oversmoothed estimate as starting point for the bootstrap procedure. Details on this issue as well as the asymptotic validity of the bootstrap method are shown in \citet{FSU}.      

Using Algorithm 3, we can obtain pointwise bootstrap confidence intervals for the nonparametric trend $g(\tau)$ with a confidence level of $\left(1-\alpha\right)$. They are denoted by $I^{(p)}_{T,\alpha}(\tau)$ and satisfy
\begin{equation}
\liminf_{n\to\infty} \mathbb{P}\left(g(\tau)\in I^{(p)}_{T,\alpha}(\tau)\right)\geq 1-\alpha, \qquad \tau\in\left(0,1\right).
\end{equation}
Using the $\alpha$-quantiles from Step 5, we compute such pointwise intervals as
\begin{equation}\label{eq:confidence}
I^{(p)}_{T,\alpha}(\tau)=\left[\hat{g}(\tau)-\hat{q}_{1-\alpha/2},\hat{g}(\tau)-\hat{q}_{\alpha/2}\right].
\end{equation}
These intervals are constructed separately for a single point $\tau$. These pointwise result are not informative about the development of the climate series over time. For this, we will compute asymptotic confidence bands satisfying
\begin{equation}
\liminf_{n\to\infty}\left[\mathbb{P}\left(g(\tau)\in I_{\alpha}(\tau) \quad \tau\in\left(0,1\right)\right)\right]=1-\alpha.
\end{equation}
Practical implementation follows a three-step procedure which was first presented in this context by \citet{Buhlmann}. It is a search algorithm based on the ordered deviations, $\hat{g}^{\ast}(\tau)-\tilde{g}(\tau)$, of bootstrapped estimates from the original estimate. The three steps are:

\begin{enumerate}
\item For all $\tau\in\left(0,1\right)$, obtain pointwise quantiles, varying $\alpha_p\in\left[1/B,\alpha\right]:\;\hat{q}_{\alpha_p/2}(\tau),\hat{q}_{1-\alpha_p/2}(\tau)$.

\item Choose $\alpha_s$ as
\begin{equation*}
\alpha_s=\operatorname{argmin}_{\alpha_p \in [1/B,\alpha]}\left|\mathbb{P}^{\ast}\left[\hat{q}_{\alpha_p/2}(\tau)\leq \hat{g}^{\ast}(\tau)-\tilde{g}(\tau)\leq \hat{q}_{1-\alpha_p/2}(\tau)\quad \forall \tau \in\left(0,1\right)\right]-(1-\alpha)\right|
\end{equation*}

\item Construct the simultaneous confidence bands as \begin{equation*}I_{\alpha_{s}}(\tau)=\left[\hat{g}(\tau)-\hat{q}_{1-\alpha_{s}/2}(\tau),\hat{g}(\tau)-\hat{q}_{\alpha_{s}/2}(\tau)\right]\quad \tau \in\left(0,1\right).
\end{equation*}
\end{enumerate}
The first step is to construct pointwise quantiles in the same way as described in Step 5 of Algorithm 3. In the second step of this procedure, a pointwise error $\alpha_s$ is found for which a fraction of approximately $\left(1-\alpha\right)$ of all centered bootstrap estimates falls completely within the resulting confidence intervals. We stress that the coverage needs to be seen over the whole sample and not point-by-point. As soon as an estimated bootstrap deviation falls outside the given intervals at one point, it is not counted for the probability in Step 2. This value $\alpha_s$ is then fixed and the resulting pointwise confidence intervals with coverage $\left(1-\alpha_s\right)$ become simultaneous confidence bands with coverage $\left(1-\alpha\right)$.

\subsection{Bootstrap algorithms for Section \ref{sec:trend_shape}}
\begin{algorithm}[Autoregressive Wild Bootstrap - Minimum location]
$\phantom{4}$
\begin{enumerate}
\item Repeat steps 1 to 4 of Algorithm 3.
\item Determine all local minima of $\hat{g}^{\ast}(t/T)$, $t=1,...,T$, and select the one closest to $t_{min}$. Denote the selected position by $t_{min}^{\ast}$.

\item Repeat Steps 1 and 2 $B$ times to obtain the empirical distribution of $t_{min}^{\ast}$.
\end{enumerate} 
\end{algorithm} 
From the $B$ locations we can now construct a $(1-\alpha)$\% confidence interval around $t_{min}$ by selecting the corresponding quantiles. In the above algorithm, we need to ensure that we identify the minimum in the bootstrap trend which corresponds to the original global minimum $t_{min}$. This does not necessarily have to be the global minimum of the bootstrap trend, which could lie far away from the original global minimum. As an empirically satisfactory solution, we therefore use the closest local minimum in Step 2. 
\begin{algorithm}[Autoregressive Wild Bootstrap - Test for a specific trend shape]
$\phantom{5}$
\begin{enumerate}
\item Estimate $\hat{g}(t/T)$ as in \eqref{eq:lc_estimator} for $t=1,...,T$. Obtain the estimate $\widehat{\boldsymbol{\theta}}$ using all data points $t\in\mathcal{G}_m$. Then define $\tilde{g}(t)$ as 
\begin{equation*}
\tilde{g}(t)\equiv\begin{cases} g_0\left(\widehat{\boldsymbol{\theta}},t\right)& \text{for }t\in\mathcal{G}_m,\\ \hat{g}(t/T)& \text{otherwise }.\end{cases}
\end{equation*} Obtain a combined residual series $\hat{u}_t=M_t\left(\hat{\epsilon}_t-\tilde{g}(t)\right)$ for $t=1,...,T$.

\item For $0 < \gamma < 1$, generate $\nu_1^\ast,\ldots,\nu_n^\ast$ as i.i.d.~$\mathcal{N}(0,1-\gamma^2)$ and let $\xi_t^* = \gamma \xi_{t-1}^* + \nu_t^*$ for $t=2,\ldots,T$. Take $\xi_1^* \sim \mathcal{N}(0,1)$ to ensure stationarity of $\{\xi_t^*\}$.

\item Calculate the bootstrap errors $u^{\ast}_t$ as $u_t^\ast=M_t\xi_t^\ast\hat{u}_t$ and generate the bootstrap observations by 
\begin{equation*}
\hat{\epsilon}_t^{\ast}=M_t\left(\tilde{g}(t)+u^{\ast}_t\right).
\end{equation*} 

\item Construct bootstrap versions of the pointwise and summary test statistics and denote them by $Q_t^{\ast}$ and $Q_i^{\ast}$ with $i=ave,sup$ for $t\in\mathcal{G}_m$.

\item Repeat Steps 2 to 4 of this algorithm $B$ times to obtain the empirical distribution of $Q_i^{\ast}$ with $i=ave,sup$ and calculate the corresponding critical values and $p$-values from it.
\end{enumerate} 
\end{algorithm}

\section{Monte Carlo study} 
We generate synthetic data to assess the finite sample performance of the methods. We focus on: (1) the size and power of the break test, (2) the confidence intervals for the break date ($T_1$) in the broken trend model, (3) the bootstrap-based test for linearity, and (4) the bootstrap-based tests for monotonicity. For extensive simulation results on the nonparametric estimator we refer the reader to \cite{FSU}. The error term process $\{u_t\}$ and the mechanism for generating the observed/missing indicators $\{M_t\}$ are adopted from the aforementioned paper. \cite{FSU} also explain how this simulation setting mimics the behavior of the Jungfraujoch data.

The process $\{u_t\}$ is generated as follows. First, we generate the error terms $\{\eta_t\}$ from an ARMA(1,1) process:
$$
 \eta_t = \phi \eta_{t-1} + \psi \epsilon_{t-1} + \epsilon_t,
 \qquad\qquad
 \epsilon_t \stackrel{i.i.d.}{\sim} 
 \mathcal{N}\left(0,\frac{(1-\phi^2) \sigma_\eta^2}{2(1+\psi^2+ 2 \phi \psi)} \right).
$$
The AR and MA parameters, respectively $\phi$ and $\psi$, are used to vary time dependence. Depending on the values of these coefficients, we adjust the variance of the i.i.d. sequence $\{\epsilon_t\}$ to keep the unconditional variance of $\{\eta_t\}$ equal to $\sigma_\eta^2/2$. Finally, we multiply these ARMA(1,1) errors by the volatility process, $u_t=\sigma_t \eta_t$. The homoskedastic scenario takes $\sigma_t=1$, whereas $\sigma_t=\sigma\left(\frac{t}{T}\right)$ with
$$
 \sigma(\tau) = \sigma_0 + (\sigma_* - \sigma_0)\tau + a \cos\left(2 \pi k \tau \right),
$$
will be used when including heteroskedasticity. We set $\sigma_0=1$, $\sigma_*=2$, $a=0.5$, and $k=4$.

Finally, we use a first order Markov chains with transition matrix
$$
 \begin{blockarray}{ccc}
& M_t=0 & M_t=1  \\
\begin{block}{c(c c)}
  M_{t-1}=0 & 0.55 & 0.45  \\
  M_{t-1}=1 & 0.2 & 0.8 \\
\end{block}
\end{blockarray}
$$
to generate the binary variables $\{M_t\}$. The stationary distribution is $\left(\frac{4}{13},\frac{9}{13}\right)$ thus indicating that approximately $\frac{9}{13}\approx 70\%$ of the sample will be classified as observed. Alternatively, a design with about 30\% of observed values is generated using $\left(\begin{smallmatrix} 0.8 & 0.2 \\ 0.45 & 0.55 \end{smallmatrix}\right)$ as the transition matrix. We consider three combinations of sample sizes and missing patterns. First, we set $T=285$ with 30\% missing observations and $T=666$ with 70\% missing observations to create two designs with approximately $200$ observations. An additional setting with $T=666$ and 30\% missing values is used to study the effects of an increasing number of data points.\footnote{The results for $T=666$ and 30\% missing observations are not reported for the monotonicity test as this is computationally too involved for the scope of this simulation study.} This gives us approximately $460$ observations. All simulation outcomes are based on $1000$ Monte Carlo replicates and $B=999$ bootstrap samples. 

\subsection{MC results for the break point test}
The data generating process for this set of simulations is:
\begin{equation}
    y_t = 4000 -0.5 T + \delta D_{t,T_1} + u_t,
\label{eq:simulatedmodel}
\end{equation}
where $T_1=0.6 T$ and we vary the value for $\delta$. The empirical rejection frequencies are reported in Panel A of Table \ref{tab:simu}. For $\delta=0$, the empirical size is always above the 5\% nominal size. The largest size distortion is observed for $T=285$ with 30\% missing and for both autoregressive and heteroskedastic errors. However, results improve when the sample size is increased to $T=666$. For $\delta>0$, the empirical power behaves as expected. That is, the test will reject more often as we move farther away from the null hypothesis (i.e. increase $\delta$) or enlarge the sample.   

\subsection{MC results for the confidence interval of the break date}
We simulate data according to \eqref{eq:simulatedmodel} with $\delta=1$. The simulated data thereby resembles the estimated parametric trend shape in the Jungfraujoch data. The empirical coverage and mean interval length are reported in panel B of Table \ref{tab:simu}. We make two observations. First, the empirical coverage is systematically below the 95\% nominal level. Similarly to the empirical size of the break point test, deviations from nominal levels are most pronounced in the presence of: (1) serial correlation, (2) heteroskedasticity, and/or (3) a high percentage of missing observations. Fortunately, for all designs, the empirical coverage moves closer towards the 95\% level when the sample size is increased. Second, the simulations also indicate that especially time-varying heteroskedasticity has a strong effect on confidence interval width.

\subsection{MC results for the linearity test}
To investigate the size of the linearity test, we generate a linear trend model without break as in  \eqref{eq:simulatedmodel} with $\delta=0$. For a study of the power of the test, we use a non-linear trend specification. It is generated, similar as in \citet{FSU}, by a smooth transition model with the parameters chosen in such a way that the resulting trend resembles the final part of the nonparametrically estimated trend in the Jungfraujoch data: an upward trend, followed by a peak and a short downward trending part. The smooth transition trend is generated as
\begin{equation*}
g(\tau) = \gamma_1 + \gamma_2 G(\tau,\lambda,c_1) +\gamma_3 G(\tau,\lambda,c_2),
\end{equation*}
with $\gamma_1=\gamma_2=1$ and $\gamma_3=-0.5$. For $\lambda>0$,
\begin{equation*}
G(\tau,\lambda,c)=\left(1+\exp\left\{-\lambda(\tau-c)\right\}\right)^{-1}
\end{equation*}
is the transition function with time as transition variable. Its inputs are time, the location of the shifts as fraction of the sample -- the parameters $c_1$ and $c_2$ -- as well as the smoothness of the shift, determined by $\lambda$. The choices $c_1=0.2$, $c_2=0.6$ and $\lambda=10$ result in the generated trend function plotted in Figure \ref{fig:smooth_trans_trend}. The error terms are generated as above with $\sigma_t=1$ as well as $\phi=0.1$ and $\psi=0$. We study three different bandwidth values ($h=0.04$, $h=0.06$, $h=0.08$). 

The results are summarized in Panel C of Table \ref{tab:simu}. The empirical size fluctuates around the nominal size of 5\%. Using the $Q_{ave}$ statistic, the test is slightly oversized in all scenarios and most accurate for the case with 70\% missing observations. The empirical power of this version is high. Using the supremum rather than the average over the pointwise test statistics results in $Q_{sup}$, for which the size is always lower compared to the other version. It is close to the nominal size in most cases and both, the size and power, increase with the bandwidth. The power of the test is low when we consider few observations; it substantially increases and reaches 1.000 when we increase the sample size.

\subsection{MC results for the monotonicity tests}
\begin{figure}
	\centering
		\includegraphics[width=0.7\linewidth, clip, trim = {0 1.5cm 0 1cm}]{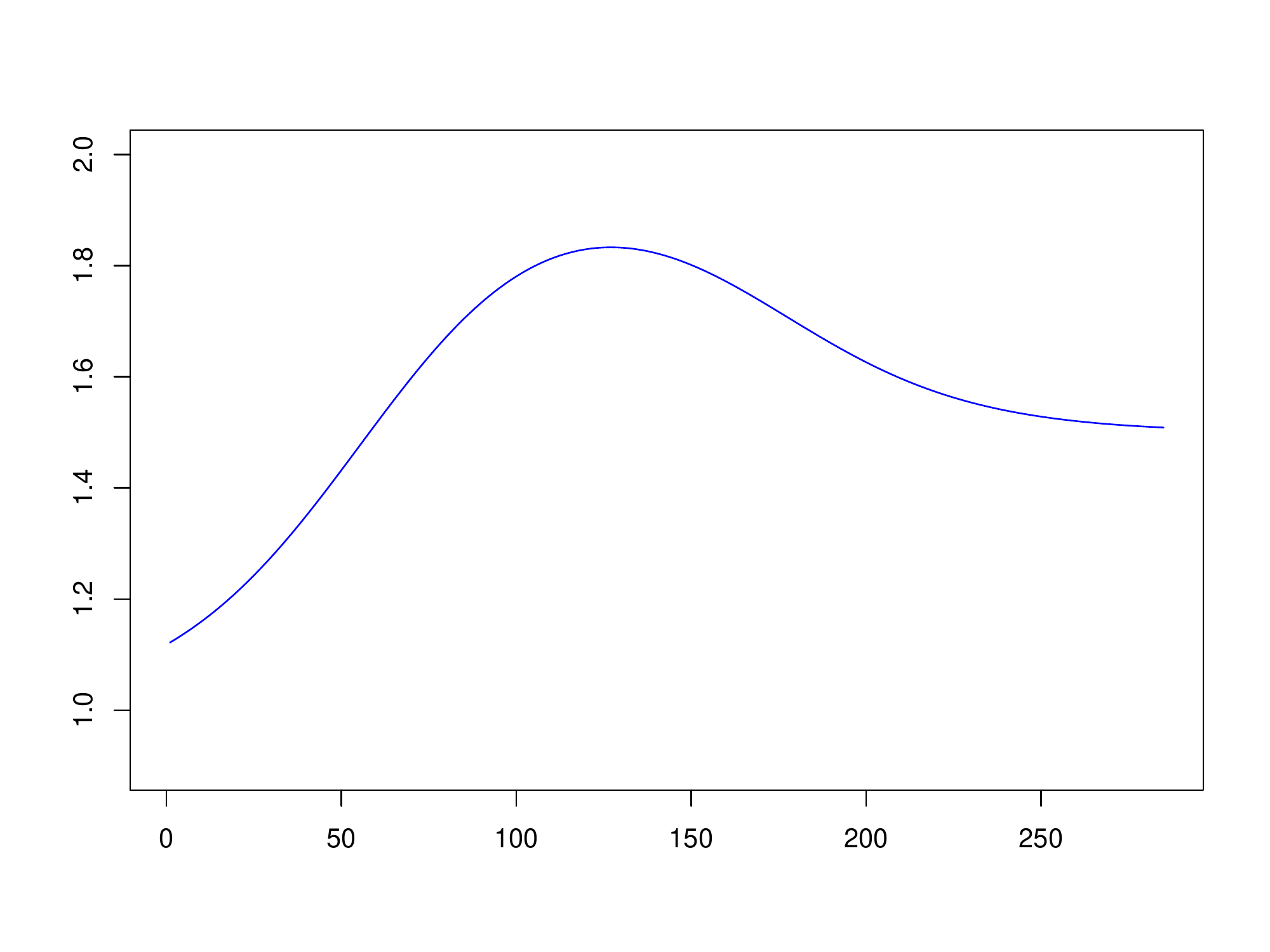}		
		\caption{Trend function generated by the smooth transition model, used to investigate the power of the linearity and monotonicity test.}
			\label{fig:smooth_trans_trend}
\end{figure}

For the monotonicity test we use exactly the same specifications as for the linearity test outlined above. To investigate size, we make use of the linear trend model without break (\eqref{eq:simulatedmodel} with $\delta=0$). Since the trend slope is positive ($\beta=0.5$), this clearly satisfies the null hypothesis of a monotonically increasing trend. For power, we simulate data from the smooth transition model. As shown in Figure \ref{fig:smooth_trans_trend}, the trend function is only monotonically increasing in the first part of the sample and then monotonically decreasing.

The results are summarized in Panel D of Table \ref{tab:simu}. The empirical size is slightly above the nominal level of 5\% for all cases. For $T=666$, both versions of the test display a size of around 7\% which is close to the nominal level. The power of the test ranges from around 68\% to 75\%, while being higher for $U_1$ than for $U_2$. Overall, both versions of the test perform similarly. 

\begin{table}[h]
\centering
\begin{tabular}{c c | c c c c c c}
\noalign{\smallskip} \hline \hline 
\multicolumn{8}{c}{\textbf{A: Break point test}} \\
\noalign{\smallskip} \hline
 & & \multicolumn{2}{c}{$T=285$} & \multicolumn{2}{c}{$T=666$} & \multicolumn{2}{c}{$T=666$} \\
 & & \multicolumn{2}{c}{30\% missing} & \multicolumn{2}{c}{70\% missing} & \multicolumn{2}{c}{30\% missing} \\
 $\phi$ & $\psi$ & $\sigma_t=1$   & $\sigma_t=\sigma(t/T)$ & $\sigma_t=1$ & $\sigma_t=\sigma(t/T)$ & $\sigma_t=1$ & $\sigma_t=\sigma(t/T)$ \\ \hline
0   & 0     & 0.088             & 0.134             & 0.101     & 0.125     & 0.076     & 0.113\\
            && (0.149)          & (0.153)           & (0.385)   & (0.213)   & (0.679)   & (0.305)\\
            && [0.345]          & [0.192]           & [0.911]   & [0.479]   & [0.999]   & [0.801]\\
0.5   & 0   & 0.176             & 0.197             & 0.107     & 0.143     & 0.107     & 0.132\\
            && (0.187)          & (0.187)           & (0.325)   & (0.218)   & (0.462)   & (0.251)\\
            && [0.277]          & [0.246]           & [0.824]   & [0.427]   & [0.934]   & [0.542]\\
0   & 0.5   & 0.138             & 0.162             & 0.077     & 0.126     & 0.107     & 0.122\\
            && (0.154)          & (0.182)           & (0.336)   & (0.217)   & (0.558)   & (0.264)\\ 
            && [0.310]          & [0.211]           & [0.824]   & [0.427]   & [0.982]   & [0.609]\\ 
\noalign{\smallskip} \hline
\multicolumn{8}{c}{\textbf{B: Confidence intervals for break location}} \\
\noalign{\smallskip} \hline
 & & \multicolumn{2}{c}{$T=285$} & \multicolumn{2}{c}{$T=666$} & \multicolumn{2}{c}{$T=666$} \\
 & & \multicolumn{2}{c}{30\% missing} & \multicolumn{2}{c}{70\% missing} & \multicolumn{2}{c}{30\% missing} \\
 $\phi$ & $\psi$ & $\sigma_t=1$   & $\sigma_t=\sigma(t/T)$ & $\sigma_t=1$ & $\sigma_t=\sigma(t/T)$ & $\sigma_t=1$ & $\sigma_t=\sigma(t/T)$ \\ \hline
 0  & 0     & 0.930     & 0.897     & 0.860     & 0.889         & 0.935     & 0.937\\
            && (20.95)  & (41.22)   & (19.79)   & (33.75)       & (13.03)   & (21.50)\\
 0.5 & 0    & 0.866     &0.798      & 0.876     & 0.903         & 0.900     & 0.927\\
            && (31.22)  & (62.52)   & (25.50)   & (45.18)       & (18.85)   & (32.30)\\
 0  & 0.5   & 0.894     & 0.849     & 0.869     & 0.919         & 0.928     & 0.934\\
            && (26.85)  & (53.18)   & (23.20)   & (40.29)       & (16.17)   & (27.13)\\ 
\noalign{\smallskip} \hline
\multicolumn{8 }{c}{\textbf{C: Linearity test}} \\
\noalign{\smallskip} \hline
& & \multicolumn{2}{c}{$T=285$} & \multicolumn{2}{c}{$T=666$} & \multicolumn{2}{c}{$T=666$} \\
 & & \multicolumn{2}{c}{30\% missing} & \multicolumn{2}{c}{70\% missing} & \multicolumn{2}{c}{30\% missing} \\
$h$ &  & $Q_{ave}$   & $Q_{sup}$ & $Q_{ave}$ & $Q_{sup}$ & $Q_{ave}$ & $Q_{sup}$\\ \hline
0.04 &   &    0.101   &   0.034   &  0.077  &  0.012 & 0.116 & 0.065 \\
&&  (0.945)  &  (0.274)  & (0.955) &  (0.207) & (1.000) & (0.935) \\ 
 0.06 &  &    0.098   &   0.047   &  0.064  &  0.043 & 0.106 & 0.082 \\
&&  (0.977)  &  (0.629)  & (0.974) &  (0.550) & (1.000) & (0.992) \\
0.08 &    &   0.088    &   0.069   &  0.066  &  0.054 & 0.088 & 0.091\\
&& (0.989)   &  (0.754)  & (0.988) &  (0.765) & (1.000) & (1.000)\\
\noalign{\smallskip} \hline
\multicolumn{8}{c}{\textbf{D: Monotonicity test}} \\
\noalign{\smallskip} \hline
& & \multicolumn{2}{c}{$T=285$} & \multicolumn{2}{c}{$T=666$} &  \\
 & & \multicolumn{2}{c}{30\% missing} & \multicolumn{2}{c}{70\% missing} \\
$h$ &  & $U_{1}$ & $U_{2}$ & $U_{1}$ & $U_{2}$ \\\hline
0.04 &   &   0.092    &   0.105   &  0.067 &  0.075 \\
                               &&  (0.685)  &  (0.754)  & (0.592) & (0.679) \\
0.06  &    &   0.095   &   0.104   &  0.069  &  0.071 \\
                               && (0.683)   &  (0.743) & (0.585) &  (0.675) \\
0.08  &    &   0.094   &   0.102   &  0.070  &  0.071 \\
                               &&   (0.687)   & (0.737) & (0.585) &  (0.667) \\ \hline
\end{tabular}
\caption{Monte Carlo simulations results. (A) Empirical rejection frequencies for the break point test. Table entries without brackets report empirical size. Numbers in parentheses (square brackets) specify empirical power for $\delta=0.05$ ($\delta=0.1$). (B) Empirical coverage probability of the break data confidence intervals (mean interval length). (C) Empirical rejection frequencies for the linearity test. Table entries without brackets report empirical size. The number between brackets is the empirical power. (D) Empirical rejection frequencies for the linearity test. Table entries without brackets report empirical size. The number between brackets is the empirical power.}
\label{tab:simu}
\end{table} 
\end{appendices}

\FloatBarrier

\end{document}